\documentclass[final,3p,twocolumn]{elsarticle}

\usepackage[subpreambles=false]{standalone}
\usepackage[utf8]{inputenc}
\usepackage[T1]{fontenc}
\usepackage{cmap}
\usepackage{graphicx}
\usepackage[dvipsnames]{xcolor}
\usepackage[hyperfootnotes=false,breaklinks=true]{hyperref}
\usepackage{setspace}
\hypersetup{
 bookmarksopen=true,
 bookmarksopenlevel=1,
 colorlinks=true,
 linkcolor=blue,
 anchorcolor=blue,
 citecolor=blue,
 filecolor=blue,
 urlcolor=blue,
 pdfpagemode=UseOutlines,
 pdfstartview={XYZ null null 1},
 linktocpage=true,
}
\usepackage{amsmath}
\usepackage{amssymb}
\usepackage{amstext}
\usepackage{amsfonts}
\usepackage{amsthm}
\usepackage{mathtools}

\usepackage{bm}
\usepackage{braket}
\usepackage{physics}
\usepackage{enumitem}
\usepackage{multirow}
\usepackage{braket}
\usepackage{tablefootnote}

\usepackage[normalem]{ulem}

\usepackage{url}

\newcommand{\changes}[1]{\textcolor{black}{#1}}

\usepackage{subfiles}

\journal{Future Generation Computer Systems}

\usepackage{eso-pic}
\AddToShipoutPictureFG*{
  \AtPageLowerLeft{%
    \raisebox{\dimexpr+0.65in}{\hspace*{0.7in}%
      \scalebox{1}{\color{gray}
        \parbox{\textwidth}{\footnotesize 
         This manuscript has been authored in part by UT-Battelle, LLC, under contract DE-AC05-00OR22725 with the US Department of Energy (DOE). The US government retains and the publisher, by accepting the work for publication, acknowledges that the US government retains a non-exclusive, paid-up, irrevocable, world-wide license to publish or reproduce the submitted manuscript version of this work, or allow others to do so, for US government purposes. DOE will provide public access to these results of federally sponsored research in accordance with the DOE Public Access Plan (\url{https://www.energy.gov/doe-public-access-plan}).
        }%
      }%
    }%
  }%
}%

\begin{document}

\begin{frontmatter}



\title{The Role of Quantum Computing in Advancing Scientific High-Performance Computing: A perspective from the ADAC Institute}
%
%

\affiliation[aff1]{organization={Oak Ridge National Laboratory},
            city={Oak Ridge},
            state={TN},
            country={USA}}

\affiliation[aff5]{organization={CSC - IT Center for Science},
            city={Espoo},
            country={Finland}}

\affiliation[aff6]{organization={Swiss National Supercomputing Centre (CSCS), ETH Zurich},
            city={Zurich},
            country={Switzerland}}

\affiliation[aff7]{organization={Pacific Northwest National Laboratory},
            city={Richland},
            state={WA},
            country={USA}}

\affiliation[aff8]{organization={National Institute of Advanced Industrial Science and Technology (AIST)},
            city={Tokyo},
            country={Japan}}

\affiliation[aff9]{organization={Centre for Development of Advanced Computing (C-DAC))},
            city={Pune},
            country={India}}

\affiliation[aff10]{organization={RIKEN Center for Computational Science},
            city={Kobe},
            country={Japan}}

\affiliation[aff11]{organization={Lawrence Berkeley National Laboratory},
            city={Berkeley},
            state={CA},
            country={USA}}

\affiliation[aff12]{organization={Pawsey Supercomputing Research Centre},
            city={Kensington},
            state={WA},
            country={Australia}}

\affiliation[aff13]{organization={Information Technology Center, University of Tokyo},
            city={Tokyo},
            country={Japan}}

\cortext[cor1]{Corresponding authors. Email: buchsg@ornl.gov, mikael.johansson@csc.fi}

\author[aff1]{Gilles Buchs\corref{cor1}} 
\author[aff1]{Thomas Beck}
\author[aff1]{Ryan Bennink}
\author[aff1]{Daniel Claudino}
\author[aff1]{Andrea Delgado}
\author[aff6]{Nur Aiman Fadel}
\author[aff1]{Peter Groszkowski}
\author[aff1]{Kathleen Hamilton}
\author[aff1]{Travis Humble}
\author[aff7]{Neeraj Kumar}
\author[aff7]{Ang Li}
\author[aff1]{Phillip Lotshaw}
\author[aff5]{Olli Mukkula}
\author[aff8]{Ryousei Takano}
\author[aff9]{Amit Saxena}
\author[aff1]{In-Saeng Suh}
\author[aff10]{Miwako Tsuji}
\author[aff11]{Roel Van Beeumen}
\author[aff12]{Ugo Varetto}
\author[aff1]{Yan Wang}
\author[aff13]{Kazuya Yamazaki}
\author[aff5]{Mikael P. Johansson\corref{cor1}}


\begin{abstract}
Quantum computing (QC) has gained significant attention over the past two decades due to its potential for speeding up classically demanding tasks. This transition from an academic focus to a thriving commercial sector is reflected in substantial global investments. While advancements in qubit counts and functionalities continues at a rapid pace, current quantum systems still lack the scalability for practical applications, facing challenges such as too high error rates and limited coherence times. This perspective paper examines the relationship between QC and high-performance computing (HPC), highlighting their complementary roles in enhancing computational efficiency. It is widely acknowledged that even fully error-corrected QCs will not be suited for all computational task\changes{s}. Rather, future compute infrastructures are anticipated to employ quantum acceleration within hybrid systems that integrate HPC and QC. While QCs can enhance classical computing, traditional HPC remains essential for maximizing quantum acceleration. This integration is a priority for supercomputing centers and companies, sparking innovation to address the challenges of merging these technologies. 
\changes{The novelty of this work lies in its unique perspective, reflecting the collective insights of the Accelerated Data Analytics and Computing (ADAC) Institute, a global consortium of over 20 leading HPC centers. Recognizing the growing importance of QC, ADAC established a Quantum Computing Working Group in 2023 to foster collaboration and knowledge-sharing among its members. This paper synthesizes insights from the group’s collaborative efforts and incorporates findings from a member survey that captures shared experiences, ongoing projects, and strategic directions. By outlining the current landscape and challenges of QC integration into HPC ecosystems, this work offers HPC specialists practical and forward-looking guidance on the opportunities and implications of QC in computationally intensive endeavors.}

\end{abstract}







\end{frontmatter}

\ifSubfilesClassLoaded{%
  \tableofcontents%
  \let\tableofcontents\relax%
  \setcounter{section}{0}%
}{}
\section{Introduction}
\label{intro}

\subsection{Context and Motivation}

Fueled by anticipated quantum speedups for classically demanding tasks, quantum computing (QC) has garnered significant interest over the past two decades, with groundbreaking innovations and record-breaking milestones being achieved at an accelerating pace.

What was once a purely academic pursuit has now evolved, with a growing number of companies developing their own quantum computers for commercial use and unveiling ambitious roadmaps. \changes{If realized, these promise rapid} growth in quantum hardware capabilities. \changes{Qubits are the basic information units in QC.} The number of useful\changes{, application relevant} qubits \changes{with error rates and connectivity sufficient to run circuits deep enough for target workloads, is projected} to grow from the \changes{current estimate of around a hundred to several thousand by the end of the decade.} \changes{If these improvements in fidelity and connectivity materialize alongside higher qubit counts, they will substantially increase the utility of QC already in the near-term for select workloads.}

This shift \changes{toward commercialization and practical application} is highlighted by McKinsey’s annual quantum technology monitor, \changes{which finds that QC companies are beginning to transition from primarily R\&D-driven activity toward revenue generation. McKinsey estimates total revenues of QC companies at roughly US\$650--750~million in 2024 and expects revenues to have surpassed US\$1~billion in 2025. Looking ahead, despite the inherent uncertainty, McKinsey projects that the QC market, encompassing both revenue and external funding, could grow to approximately US\$16–37 billion by 2030. The potential economic value unlocked by QC applications is estimated at roughly US\$0.9--2.0~trillion by 2035. Investment activity also remains concentrated in QC: total start-up investment reached about US\$2.0~billion in 2024 (up roughly 50\% year over year), with more than 80\% of this investment directed toward QC~\cite{McKinseyQuantumMonitor2025}.}

Despite increasing qubit counts and functional advancements, it should be duly noted that today’s quantum computers are yet to achieve the scale needed for \changes{practical applications}. The main obstacles in addition to limited qubit counts are error rates, limited coherence times, and other engineering challenges, as will be discussed in the following sections. Although technical and scientific advances are rapid, these fundamental issues do cast some uncertainty over the near-term applicability of QC. Indeed, optimistic projections suggest that a practical quantum advantage could be achieved within a few years, while pessimistic views argue that the technical challenges may remain insurmountable for the foreseeable future. In a landscape where media coverage often highlights the most extreme predictions, it can be challenging to discern the current reality of QC.

It is widely acknowledged that even fully error-corrected quantum computers will not be suitable for all computational tasks, even if they in principle can be universal computing machines. The prevailing consensus is that they will likely serve as quantum accelerators within hybrid systems combining quantum and high-performance computing (HPC). \changes{This is conceptually similar, though fundamentally different in mechanism, to the role that Graphics Processing Units (GPU) play in enhancing performance for targeted tasks. Consequently, the practical near-term challenge lies in understanding how to effectively integrate, operate, and evaluate QC within HPC environments, focusing on aspects such as scheduling, orchestration, reproducibility, and workflow-level metrics. While quantum processing unit (QPU) may eventually accelerate select kernels, classical HPC is already essential for making quantum acceleration usable in practice. In hybrid workflows, HPC typically provides compilation and hardware-aware circuit optimization, large-scale emulation/validation, and the classical pre-/post-processing needed for statistical estimation, error mitigation, and (in fault-tolerant regimes) decoding and control services. Accordingly, a central theme of this paper is that QC and HPC must be co-designed at the workflow level to ensure any quantum speedup translates into competitive end-to-end time- and energy-to-solution, a challenge that has increasingly driven supercomputing centers and companies to innovate in integrating these fundamentally different technologies.}

\subsection{Objectives}
The purpose of this work is to introduce QC to readers with an HPC background who may have limited previous familiarity with the QC field. We provide a clear and concise overview while reflecting the collective \changes{and unique} perspective of the Accelerated Data Analytics and Computing Institute (ADAC) network~\cite{ADAC}. Founded in 2016, ADAC is dedicated to advancing HPC capabilities, particularly by preparing application software for hybrid accelerated architectures and expanding the range of applications on GPU-accelerated supercomputers. With more than 20 member organizations managing \changes{some of the world's leading HPC centers}, ADAC supports academia, government, and industry in addressing some of the world’s most significant scientific challenges.

In recognition of the growing importance of QC and the emergence of experimental efforts to integrate it with HPC systems, ADAC established a dedicated Quantum Computing Working Group during its 12th workshop in 2023. The QC working group complements four initial HPC-focused groups. Recently, a sixth group has been established that focuses on training, outreach, and workforce development. The QC working group aims to promote global collaboration and knowledge-sharing in this rapidly evolving field.

This article is a product of this \changes{unique} initiative, synthesizing insights from collaborative efforts within the working group. To improve its outlook, the paper also incorporates findings from an ADAC members' survey \changes{(see Sec.~\ref{survey})}, capturing shared experiences, ongoing projects, and strategic directions. Collectively, these elements aim to guide the HPC community in navigating the challenges and opportunities of QC integration.

\subsection{Structure of the article}
The article is structured as follows. In Sec.~\ref{QC}, we provide a high-level introduction to QC~(\ref{QC_basics}) \changes{including an overview of current quantum devices operating with limited qubit counts and noise~(\ref{NISQ}), as well as more advanced systems designed for error correction and scalable computation~(\ref{FTQC}). Additionally, we discuss the various physical platforms used to implement qubits~(\ref{platforms}).} These offer the needed foundational knowledge for non-expert readers to follow the rest of the paper. For those seeking a more in-depth understanding of QC, we recommend consulting textbooks such as~\cite{Nielsen&Chuang,YanofskyMannucci2008,Aaronson2013,Bernhardt2019,Hidary2019,Wong2022}. 

\changes{In Sec.~\ref{integration}, we discuss the integration of QC with HPC and artificial intelligence (AI), highlighting the synergy between these technologies. The section focuses on hardware integration~(\ref{HW integration}), energy efficiency ~(\ref{energy efficiency}), software integration~(\ref{software integration}) followed by a discussion on leveraging AI to enhance QC~(\ref{AI for QC}). A detailed discussion on  the relevance of HPC for real-time quantum error correction is presented in Sec.~\ref{real_time}. It concludes with a detailed comparison between the different integration models~\ref{maturity int mods}. 
\\
\indent In Sec.~\ref{use_cases}, we explore QC use cases in HPC. Sec.~\ref{use_cases_intro} introduces the concept of quantum algorithms and quantum utility. The remainder of the section focuses on domain-specific applications, including condensed matter physics~(\ref{CMP}) and high energy/nuclear physics~(\ref{HEP}), quantum chemistry~(\ref{QChem}), cryptography~(\ref{Crypto}), combinatorial optimization~(\ref{Optim}) and AI~(\ref{AI}).
\\
\indent Sec.~\ref{emulation} elaborates on the critical role of emulation in QC. It begins by defining the benefits of emulation as the practice of simulating QC systems on classical HPC infrastructures~(\ref{emulation_benefits}), and outlines its importance in the context of certifying quantum advantage~(\ref{certification}). Prevalent emulation tools and platforms, covering their capabilities and limitations when applied to current \changes{noisy} systems are then presented~(\ref{emulation methods}).
\\
\indent Sec.~\ref{benchmarking} details benchmarking strategies for quantum systems, starting with microbenchmarking for low-level hardware metrics~(\ref{microbenchmarking}), followed by synthetic~(\ref{synthetic benchmarking}) and application-specific benchmarking~(\ref{application specific benchmarking}) to evaluate workloads and real-world scenarios. Benchmarking suites~(\ref{benchmarking suites}) provide comprehensive tools, and finally, joint HPC+QC benchmarking~(\ref{joint benchmarking}) assesses the performance of hybrid quantum-classical workflows.
\\
\indent Sec.~\ref{survey} summarizes findings from an ADAC member's survey showcasing shared experiences, highlighting ongoing projects, and outlining strategic directions that reflect their perspectives on the role of QC in advancing scientific HPC. The section covers survey design~(\ref{survey design}), infrastructure~(\ref{infrastructure}), user demand~(\ref{user demand}), training~(\ref{training}), strategic planning~(\ref{strategic planning}), and a final summary~(\ref{survey summary}).
\\
\indent Finally, Sec.~\ref{conclusions} concludes by emphasizing the potential of QC to transform HPC through hybrid systems. It underscores the progress, challenges, and importance of global collaboration and innovation to advance QC+HPC integration for tackling complex scientific challenges.
} 

\ifSubfilesClassLoaded{%
  \nocite{apsrev41Control}%
  \bibliographystyle{apsrev4-1}%
  \bibliography{refs}%
}{}
\end{document}

\ifSubfilesClassLoaded{%
  \tableofcontents%
  \let\tableofcontents\relax%
  \setcounter{section}{1}%
}{}
\section{Quantum Computing}
\label{QC}

\subsection{Quantum Computing Basics}
\label{QC_basics}

In classical computing, the basic information unit is the binary bit. The key component of a quantum computer is the quantum bit, or qubit, a quantum system that can represent two well-defined quantum states generally expressed mathematically as $\ket{0}$ and $\ket{1}$, known as the computational basis states\footnote{The states $\ket{i}$ are called \textit{ket} in Dirac's notation and correspond to column vectors, for instance $\ket{0}=$ \scalebox{0.7}{$\begin{pmatrix} 1 \\ 0 \end{pmatrix}$}, $\ket{1}=$ \scalebox{0.7}{$\begin{pmatrix} 0 \\ 1 \end{pmatrix}$}, and quantum gates can be represented as matrices acting on these vectors, e.g. the Pauli-X gate acting on $\ket{0}$ writes: $X\ket{0}=$ \scalebox{0.7}{$\begin{pmatrix} 0 & 1 \\ 1 & 0 \end{pmatrix} \begin{pmatrix} 1 \\ 0 \end{pmatrix} = \begin{pmatrix} 0 \\ 1 \end{pmatrix}$}$=\ket{1}$, see e.g.~\cite{Nielsen&Chuang} for more detail.}. These states may correspond to an atom or ion's ground and excited states, a photon's horizontal and vertical polarization states, or an electron's spin-up and spin-down among other possible degrees of freedom, see Sec.~\ref{platforms}. Unlike classical bits, a qubit can exist in a \changes{coherent} superposition of $\ket{0}$ and $\ket{1}$, meaning \changes{its state is described by complex probability amplitudes until measurement}. This is due to the fundamental nature of quantum mechanics, where the state of a system, mathematically expressed by a wavefunction $\ket{\psi}$, is described by a linear combination of basis states. For example, a qubit in a superposition state can be written as:
\begin{equation} \label{eq_super}
    \ket{\psi}=\alpha \ket{0}+\beta \ket{1}
\end{equation}
where $\alpha$ and $\beta$ are complex numbers representing the probability amplitudes for the states $\ket{0}$ and $\ket{1}$, respectively. Upon measurement, the wavefunction "collapses" (within the so-called Copenhagen interpretation of quantum mechanics) onto one of the states $\ket{0}$ or $\ket{1}$ with probabilities given by the squared magnitudes of the amplitudes, i.e., $|\alpha|^2$ and $|\beta|^2$, with the constraint that $|\alpha|^2 + |\beta|^2 = 1$. This probabilistic nature of quantum states means that the outcome of a quantum measurement is not deterministic but rather follows a probability distribution based on the state before a measurement. \changes{In practice, quantum circuits are executed multiple times, referred to as “shots,” to collect measurement statistics, which are then used to estimate either the probabilities of specific measurement outcomes or the expectation values of quantum observables.}

Additionally, qubits can become entangled, one of the most remarkable features of quantum mechanics. Entanglement occurs when the quantum state of two or more qubits becomes correlated in such a way that the state of one qubit cannot be described independently of the state of the others. Even if the qubits are physically separated by large distances, for example the measurement of one qubit instantaneously affects the measurement outcome of the other, a phenomenon which Albert Einstein famously referred to as "spooky action at a distance"~\cite{bromberg1972twentieth}. \changes{It is important to note that this does not imply faster-than-light information transfer. While measurement outcomes are correlated due to entanglement, no actual information is transmitted between particles without the aid of classical communication~\cite{Nielsen&Chuang}.} Mathematically, an entangled state for two qubits can be expressed, for example, as:

\begin{equation} \label{eq_entangled}
    \ket{\psi}=\frac{1}{\sqrt{2}} (\ket{00}+\ket{11})
\end{equation}

This state, known as a Bell state, is an example in which the individual states of the qubits are undefined and the system is described as a combined state. In this specific Bell state, we have an equal superposition of the states $\ket{00}$ and $\ket{11}$. Importantly, the two other possible states, $\ket{01}$ and $\ket{10}$, are absent from the superposition. When a measurement is made on one qubit, it forces the other qubit into a corresponding state, regardless of the distance between them. The measurement of the first qubit gives either $\ket{0}$ or $\ket{1}$ as a result with equal probability, as we have an equal superposition. The measurement of the first qubit also immediately forces the value of the second qubit to the same state. Before measuring, we do not know the state of either the first or second qubit, but we \emph{do} know that they have to be equal.
\changes{The third essential ingredient of QC is interference. Quantum interference allows the amplitudes of quantum states, which are complex numbers representing probabilities (see eq. \ref{eq_super}), to combine either constructively (enhancing desired outcomes) or destructively (canceling out undesirable ones). This mechanism is crucial in quantum algorithms, as it enables the computation to focus on correct solutions by amplifying their probabilities while suppressing incorrect ones.
\\
\indent These three ingredients, i.e. i) superposition, ii) entanglement, and iii) interference, work together to power quantum algorithms. Superposition enables quantum systems to represent many possibilities simultaneously, entanglement creates correlations that allow qubits to work as a connected system, and interference guides the computation by amplifying correct solutions and suppressing incorrect ones. For example, in Grover's algorithm, the quantum superposition represents all candidate solutions, and the iterative application of the oracle and diffusion operators typically generates multi-qubit entanglement within the search register. Constructive and destructive interference then amplify the probability of measuring the correct answer by enhancing the amplitude of the marked state(s)~\cite{grover1996fast}. By leveraging these phenomena, quantum computers have the potential to surpass classical systems for certain types of problems, but the full extent of their advantages remains an area of active research (see Sec.~\ref{use_cases}).}

Logical quantum gates, which manipulate the state of qubits, are used to perform operations within quantum algorithms. These gates operate on individual qubits by changing their probabilities, enabling transformations such as creating superpositions or entangling qubits. In contrast to classical gates, which operate deterministically on classical bits, quantum gates allow for operations that take advantage of quantum phenomena like superposition and entanglement, making them capable of performing computations that are fundamentally different from, and out-of-reach of classical computers. For example, a typical one-qubit gate is the Pauli-X gate, which acts like a classical NOT gate, flipping a qubit from $\ket{0}$ to $\ket{1}$ and vice versa, and the Hadamard gate that transforms a qubit in the state $\ket{0}$ or $\ket{1}$ into a superposition similar to eq.~\ref{eq_super}. A typical two-qubit gate is the CNOT gate (Controlled-NOT) that flips the state of one (target) qubit based on the state of another (control) qubit, a key operation for creating entanglement between qubits. 

Similarly to classical computation, where universality is achieved with a simple set of gates like the NAND gate that can implement any logical operation, QC relies on a \emph{universal gate set}: \changes{a finite collection of gates from which any unitary operation (and thus any quantum algorithm, up to approximation) can be constructed.} This set typically includes gates such as the Hadamard, CNOT, and T gates. The latter is a so-called \changes{phase} shift gate that modifies the phase of a quantum state by an angle $\varphi$, for instance here $\ket{0}\rightarrow\ket{0}$ and $\ket{1}\rightarrow e^{i\frac{\pi}{4}}\ket{1}$.

The ability to apply sequences of quantum gates with high precision in quantum circuits may eventually allow quantum computers to fully harness the power of quantum mechanics in problem-solving, though this capability is not yet fully realized, as outlined further. 

While two-state qubits are presently the most common approach to implement universal quantum computers, it is also possible to use more states to implement the basic information unit in QC. General multi-level computational units are called qudits -- combining for example three states,  $\ket{0}$, $\ket{1}$, and $\ket{2}$, would give qutrits. Qudits provide an even larger state-space for processing information, but the QC approach remain similar~\cite{qudits2020}. In the following, we will use qubits for the discussion, but note that the general concepts apply equally to qudits. 

Table \ref{fig:classical_vs_quantum} summarizes the key differences between classical and QC concepts.

\begin{table*}[htbp]
\begin{tabular}{|p{4cm}|p{6cm}|p{6cm}|}
\hline
\textbf{Feature} & \textbf{Classical Computing} & \textbf{Quantum Computing} \\
\hline
Basic Unit of Information & Bit: can be 0 or 1 & Qubit: can be 0, 1, or a \emph{superposition} of both \\
\hline
Data Representation & Definite state (0 or 1) at any given time & Superposition states allow simultaneous representations of multiple states\\
\hline
Processing Mechanism & Deterministic. Follows fixed logic gates (AND, OR, NOT, ...) & \changes{Deterministic unitary evolution + probabilistic measurement.} Uses quantum gates that manipulate qubits through unitary transformations \\
\hline
State Readout & Reads bits without affecting them & Measurement collapses quantum state of the qubit \\
\hline
Key Phenomena & Boolean logic, deterministic algorithms & Superposition, entanglement, quantum interference \\
\hline
Processing Power & Scales linearly with number of bits & Can scale exponentially with number of qubits for certain \changes{algorithms by leveraging superposition, entanglement and interference as key quantum phenomena} \\
\hline
Error Susceptibility & Robust to noise; classical error correction available & Sensitive to noise and decoherence; needs quantum error correction \\
\hline
Algorithms & Classical algorithms (sorting, searching, simulation, ...) & Quantum algorithms (Shor’s, Grover’s, quantum simulation, ...) \\
\hline
Applications & General-purpose computing, office software, simulations & Optimization, quantum simulation, cryptography, ML/AI acceleration \\
\hline
Speed Advantage & Efficient for most everyday computing tasks and modelling & Potential exponential speed-up for specific problems \\
\hline
Hardware & CMOS transistors, silicon chips & Several different implementations \\
\hline
Maturity & Mature, mass-produced, widely used & Early-stage, experimental, limited qubits \\
\hline
\end{tabular}
\caption{Main differences between classical and QC concepts}
\label{fig:classical_vs_quantum}
\end{table*}

\subsection{Noisy Intermediate-Scale Quantum (NISQ) Devices}
\label{NISQ}

Unlike the digital states of classical bits, qubit states are highly sensitive to small environmental perturbations and operational inaccuracies. These noise sources can lead to decoherence, which disrupts quantum superposition and entanglement, as well as gate fidelity errors, where quantum gates fail to perform as intended. 
In this context, noisy Intermediate-Scale Quantum (NISQ) systems represent the current generation of quantum computers, characterized by their limited qubit counts and susceptibility to noise and errors during operations~\cite{Preskill2018}. 
Despite their imperfections, NISQ systems have showed remarkable progress in exploring quantum algorithms ~\cite{bharti2022noisy} across disciplines such as physics (see Sec.~\ref{CMP} and~\ref{HEP}), quantum chemistry (Sec.~\ref{QChem}), machine learning (ML) (Sec.~\ref{AI}), and combinatorial optimization (Sec.~\ref{Optim}). This progress has been achieved mostly in the form of hybrid quantum-classical algorithms, in which a classical algorithm uses a limited NISQ device to help solve a small but difficult part of a larger problem.
Variational quantum algorithms (VQA) in which the NISQ device evaluates a quantum objective function have been especially popular~\cite{bharti2022noisy}.
Remarkably, NISQ machines have recently been able to simulate physical systems (see Sec.~\ref{CMP}) at scales \changes{that are challenging for leading} classical approaches\changes{, using} specialized applications of analog devices~\cite{Daley2022,flannigan2022propagation}.
While these achievements mark substantial progress, they also underscore the limitations inherent in NISQ systems. Their noisy and error-prone nature severely constrains their scalability and accuracy, emphasizing the need for advancements in quantum error correction \changes{(QEC)} and the development of \changes{fault tolerant quantum computing} FTQC to fully realize the transformative potential of QC.

\subsection{Fault Tolerant Quantum Computing (FTQC)}
\label{FTQC}

Quantum computers can be made fault tolerant by encoding each logical qubit of an algorithm across multiple physical qubits in a manner reminiscent of that of classical error correction codes. Provided the hardware errors are small enough, the error per \emph{logical} quantum operation can be made arbitrarily small by using a sufficiently large amount of physical redundancy~\cite{roffe2019quantum}. The resource overhead needed for \changes{QEC} appears to be vastly greater than that for classical error correction, however. The overhead varies depending on qubit \changes{implementations}. Presently, trapped ion implementations appear to have the potential for the lowest ratio between physical and logical, error-resilient qubits, with an overhead of about 10-100 physical qubits per logical qubit. For other modalities, such as superconducting qubits, the ratio was for a long time expected to be counted in the hundreds, see e.g.~\cite{Fowler_2012,Beverland2022,Dalzell2023}. 

Similarly, each logical operation involves several orders of magnitude overhead in native hardware operations. Certain essential operations known as T gates are especially costly:  substantial portions of a quantum computer must be devoted to high-fidelity preparation of resource states called ``magic states'' that are consumed by T gates.

Therefore, the vast majority of a quantum computer's physical resources will be devoted to \changes{FTQC}. The development and resource optimization of fault tolerance protocols is a highly active area of research, and requires the development and use of corresponding compilation tools. Progress towards lower overhead has recently been rapid. IBM is, for example presently pursuing so-called bivariate bicycle codes, where 12 logical qubits can be encoded with as few as 144 physical qubits \cite{yoder2025tourgrossmodularquantum}. Equally important is the improvement of the error rates and speed of quantum gates, as these factors work in conjunction with \changes{QEC} improvements to enhance the overall reliability and scalability of quantum computers. In this context, recent experimental breakthroughs by companies like QuEra~\cite{Bluvstein2023}, Microsoft-Quantinuum~\cite{tesseract_2024}, Microsoft-Atom Computing~\cite{MS-AC_2025} and Google~\cite{acharya2024quantum}, leveraging neutral atoms, trapped ions, and superconducting qubits, respectively, represent particularly encouraging advancements in the field. Notably, while the first two efforts demonstrated logical qubits with reduced error rates, Google's work further showcased exponential suppression of logical errors as the size of the physical qubit array increased, a critical milestone for scalable QC.

Parallel improvements in both hardware and software have the potential to significantly squeeze the expected time-frame for reaching practical utility for quantum-accelerated HPC. As a prominent example, breaking RSA-2048 encryption was for a long time expected to require on the order of 20 million physical qubits. In May 2025, Gidney lowered the estimate by 95\%, that is, showed how to factor 2048 bit RSA integers with less than a million noisy, \changes{physical} qubits \changes{at assumed error rates and error-correction scheme}, corresponding to roughly 1,600 logical qubits \cite{gidney2025factor2048bitrsa}.

\subsection{Qubit Platforms}
\label{platforms}

The physical implementation of a physical qubit does not have to be an ideal two-state system; it only needs to behave as one in a well-controlled environment. This flexibility has enabled the development of various QC platforms, each coming with its own set of advantages and challenges. Some technologies are more mature than others, and some are more suitable for near-term, noisy, small-scale quantum computers. It remains too early to determine which technology holds the most potential in the long term, as significant advancements are still needed before quantum computers can achieve the accuracy and scale required for practical, real-world modeling applications.

Current QC systems are based on qubit platforms including:

\begin{description}
    \item \textit{Superconducting qubits} implemented as superconducting resonant circuits that are operated in extremely low temperatures~\cite{superconducting_2019,superconducting_2020}.
    \item \textit{Trapped ion qubits} implemented using electromagnetically confined and laser-cooled charged atomic particles~\cite{trapped_ions_2019}.
    \item \textit{Neutral atom qubits} implemented using electromagnetically confined and laser-cooled Rydberg atoms~\cite{Neutral_2020}. 
    \item \textit{Photonic qubits} implemented by manipulating optical properties of light~\cite{photonic_review_short_2019,photonic_review_long_2018}.
    \item \textit{Semiconductor spin qubits} implemented by manipulating the spin of individual charge carriers in semiconductor structures such as quantum dots or dopants arrays~\cite{Quantum_Dots_RMP_2023,Zwanenburg_RMP_2013,hu2025single}.
    \item \textit{Nitrogen-vacancy centers} leveraging spin states of electrons in diamond lattice defects~\cite{NV_2021}. 
    \item \textit{Topological qubits}, based on topological properties of quasiparticles called Majorana fermions~\cite{Topological_2008,Topological_2022}. 
\end{description}

\changes{Variations in QC modalities significantly impact their integration with HPC systems. Factors such as gate times, coherence properties, and control requirements determine latency budgets, data rates, and facility constraints, meaning that the ideal integration and orchestration approach is closely tied to the specific qubit modality. We return to these implications in Sec.~\ref{integration} (Secs.~\ref{HW integration},~\ref{software integration}, and~\ref{energy efficiency}).} 

The main properties of these different qubit types are listed in table \ref{fig:hardware_landscape}. A major distinction between these technologies is that some leverage quantum systems found in nature, while others are artificial constructs designed to simulate two-state quantum systems.

Natural quantum systems, such as spin in atomic particles, benefit from identical individual qubits. In comparison, artificial qubits are susceptible to fabrication defects, which can make controlling qubits more difficult, but benefit from more flexibility in design and fabrication.

Superconducting qubits are currently very popular because they have demonstrated a good combination of attributes, most notably fast and high-fidelity 2-qubit gates, and relatively good scaling properties, at least for NISQ systems. One of the biggest challenges with superconducting qubits is their short coherence times.

Trapped ions~\cite{trapped_ions_2019} and neutral atoms~\cite{Neutral_2020} also offer significant advantages for QC, with trapped ions providing all-to-all qubit connectivity, reducing the need for costly SWAP gates, and high gate fidelities. Both platforms feature long coherence times, far surpassing those of superconducting qubits, but feature slower gate speeds and scaling challenges. Neutral atoms in optical tweezers have recently made significant advances using dynamically reconfigurable connectivity~\cite{Bluvstein2023}.

Photonics QC~\cite{photonic_review_long_2018,photonic_review_short_2019} currently involves two main approaches. The first is Gaussian Boson Sampling, a non-universal type of QC designed for specific problems. The second approach is universal photonic QC, which has faced challenges, particularly in executing two-qubit gates using photons. Despite these hurdles, advancements are being made, including techniques like measurement-based QC. Key advantages are room temperature operation and the potential for very fast quantum gates.

Semiconductor spin qubits~\cite{Quantum_Dots_RMP_2023,Zwanenburg_RMP_2013} and NV centers~\cite{NV_2021} are both emerging qubit technologies that are currently less developed than their counterparts, so far only demonstrated with a low qubit count. Semiconductor spin qubits have shown fast gates on par with superconducting qubits, and could potentially leverage modern microelectronics fabrication techniques. NV centers can achieve high gate fidelities and operate without cryogenic cooling, though scalability remains a challenge.

Topological qubits\footnote{Microsoft recently announced their Majorana 1 chip~\cite{microsoft2025interferometric}, based on the so-called "topoconductor" (or topological superconductor) materials, "\emph{to observe and control Majorana particles}"~\cite{mzm}. However, their formal publication admitted that the current measurements by themselves cannot determine if the low-energy states detected are topological~\cite{microsoft2025interferometric}. Therefore, there are still concerns about the existence of topological qubits~\cite{castelvecchi2025microsoft}.} promise better error rates than their counterparts but are still in a fundamental research phase~\cite{Topological_2022,microsoft2025interferometric}. 

It is worth mentioning that for most qubit realizations, technological constraints limit the size of current \changes{QPUs}. Because the power of QC grows with the number of coherent qubits (while maintaining low error rates), such limitations prevent them from achieving a quantum advantage. \changes{One promising path to scalability is modular or distributed quantum computing (DQC)~\cite{barral2024_DQC}, which employs quantum interconnects across local or wide-area quantum networks. By leveraging the unique strengths and mitigating the weaknesses of various qubit technologies, DQC can enable an integrated approach that optimally combines the best features of each platform.} Although recent DQC breakthroughs have been demonstrated with identical QPUs, for instance trapped ions~\cite{main2025distributed}, substantial research is still required to understand how to coherently and efficiently transfer quantum information between different qubit modalities~\cite{lauk2020perspectives, Martinis_2025}.
\newline

\begin{table*}[htbp]
\footnotesize
\begin{tabular}{ |p{2.1cm}|p{2.1cm}|p{2.1cm}|p{2.1cm}|p{2.1cm}|p{2.1cm}|p{2.1cm}|} 
 \hline
 & \begin{center} Superconducting qubits \end{center}
 & \begin{center} Trapped ions \end{center}
 & \begin{center} Neutral atoms \end{center}
 & \begin{center} Photonics \end{center}
 & \begin{center} Semiconductor spin qubits \end{center}
 & \begin{center} Nitrogen-Vacancy centers \end{center} \\
 \hline 

 \begin{center} Gate fidelities \\ (2-qubit gates) \end{center}
 & \begin{center} \changes{$\sim 99.9$ \%} \end{center}
 & \begin{center} $ >99.9$ \% \end{center}
 & \begin{center} $ >99.5$ \% \end{center}
 & \begin{center} $ >99.0$ \% \end{center}
 & \begin{center} $> 99.5$ \% \end{center} 
 & \begin{center} $\sim 99.9$ \% \end{center} \\
 \hline

 \begin{center} Coherence times\textsuperscript{1} \end{center}
 & \begin{center} From hundreds of microseconds to millisecond  \end{center} 
 & \begin{center} From multiple seconds ($T_2$) to minutes ($T_1$)  \end{center} 
 & \begin{center} Up to multiple seconds ($T_1$)  \end{center} 
 & \begin{center} Not always applicable\textsuperscript{2} \end{center} 
 & \begin{center} From tens to hundreds of microseconds  \end{center}
 & \begin{center} Hundreds of milliseconds  \end{center} \\
 \hline

  \begin{center} Gate lengths \end{center}
 & \begin{center} From tens to hundreds of nanoseconds \end{center} 
 & \begin{center} Hundreds of microseconds  \end{center}
 & \begin{center} Microseconds  \end{center}
 & \begin{center} Tens of nanoseconds\textsuperscript{4}  \end{center}
 & \begin{center} Hundreds of nanoseconds  \end{center}
 & \begin{center} Microseconds  \end{center} \\
 \hline

 \begin{center} Connectivity \end{center}
 & \begin{center} Nearest neighbor \end{center} 
 & \begin{center} All-to-all \end{center} 
 & \begin{center} All-to-all \end{center} 
 & \begin{center} Not always applicable\textsuperscript{2} \end{center}  
 & \begin{center} Nearest neighbor \end{center}
 & \begin{center} Nearest neighbor \end{center} \\
 \hline

 \begin{center} Physical qubits \textsuperscript{3} \end{center}
 & \begin{center} \changes{$\sim 1000$, $\sim 150$ qubits in production systems} \end{center}
 & \begin{center} \changes{$\sim 100$} qubits  \end{center}
 & \begin{center} \changes{$\sim 1000$} qubits \end{center}
 & \begin{center} $\sim$ 10 with universal \\ 100+ with GBS\end{center}
 & \begin{center} $\sim$ 10 qubits \end{center}
 & \begin{center} $\sim$ 10 qubits \end{center} \\
 \hline

\begin{center} Summary \end{center}
 & \begin{center} Fast and high-fidelity gates. Good scaling properties but limited qubit connectivity. Challenges with short coherence times. \end{center}
 & \begin{center} Longest coherence times and highest gate fidelities, as well as all-to-all connectivity. Challenges with scaling and slow gates. \end{center}
 & \begin{center} Long coherence times and medium operation speed. Very good scaling with all-to-all connectivity. Challenges with fidelities. \end{center}
 & \begin{center} Room temperature operation and very fast gate operations. Universal two-qubit operations are challenging to implement. \end{center}
& \begin{center} Potentially well-scaling technology synergizing well with silicon fabrication methods. Lack of demonstrations for scaling.\end{center}
& \begin{center} Potentially high gate fidelities and long coherence times at room temperature. Lack of demonstrations for scaling. \end{center} \\
\hline
\end{tabular}
\caption{
Table of quality metrics \changes{provisional as of January 2026} for different qubit modalities. These numbers are estimates for state-of-the-art commercial QC platforms. They should not be interpreted as the exact record benchmarked results for individual devices. \changes{As vendors report these metrics using different protocols, the entries should be interpreted as order-of-magnitude indicators rather than directly comparable benchmarks.} Semiconductor spin qubits and NV-centers are emerging technologies, with limited commercial use, thus quality metric estimates may be optimistic and not directly comparable to other qubit modalities. Topological qubits are omitted from the table due to the lack of data. \changes{Estimates for quality metrics are based on following sources: 
Superconducting qubits~\cite{IBM2025BlogNighthawk, IBM2026Processors, IBMQ2024Resources, IQM2024benchmark, IQM2024milestone, IQM2026Products}, 
Trapped ions~\cite{Quantinuum2026Helios, IonQ2026Systems, chen2023benchmarkingtrappedionquantumcomputer}, 
Neutral atoms~\cite{QuEra2026Gemini, AtomComputing2026AC1000, AtomComputing2026AC1000Fidelitypaper, Novonordisk2025Magne, QuEra2023Aquila, Wintersperger2023NeutralAQ, Xia_2015NeutralAtomSingleQubitFidelity}, 
Photonics~\cite{PsiQuantum2024Blueprint, Psiquantum2025manufacture}, 
Semiconductor spin qubits~\cite{Neyens2024ProbingSpinQubits, Noiri2022FastUniversalGatesSilicon}
Nitrogen vacancy centers~\cite{joas2024highfidelityDiamondSpinGates, bartling2024universalhighfidelityquantumgates}
}
\newline
\textsuperscript{1} Coherence times are usually reported for two different depolarization processes, $T_1$ and $T_2$. Some qubit modalities have their coherence times bounded by $T_2$ \\
\textsuperscript{2} Due to the difficulty of entangling photons, photonic QC is often implemented with Gaussian Boson Sampling (GBS) or measurement-based QC techniques. Coherence times and connectivity are not applicable metric for these techniques. \\
\textsuperscript{3} While some qubit modalities have demonstrated high scalability with 1000+ qubit counts, these systems most likely had significantly lower qubit quality than current state-of-the-art commercial QPUs. \\
\textsuperscript{4} The low nanosecond timescale reflects the intrinsic photonic processes that occur at nearly the speed of light or the operational speed of fast modulators and detectors. However, this does not account for delays in feedforward control, state preparation, or error correction, which operate on much slower timescales.
}
\label{fig:hardware_landscape}
\end{table*}

\ifSubfilesClassLoaded{%
  \nocite{apsrev41Control}%
  \bibliographystyle{apsrev4-1}%
  \bibliography{refs}%
}{}

\end{document}

\ifSubfilesClassLoaded{%
  \tableofcontents%
  \let\tableofcontents\relax%
  \setcounter{section}{2}%
}{}
\section{Integration of Quantum Computing, HPC, and AI}
\label{integration}

The idea of using quantum computers as accelerators dates back several decades. For example, in Shor’s factoring algorithm~\cite{shor1994algorithms}, a quantum computer efficiently solves the classically intractable period-finding problem, with the results subsequently processed by a classical computer to complete the factoring task. Here, the classical part forms a rather small part of the workflow. More recently, this concept has been extended to quantum-classical hybrid computing~\cite{bharti2022noisy}, as seen, for example, in Variational Quantum Algorithms (VQAs) (see Sect.~\ref{NISQ}), where classical devices handle optimization tasks and work in tandem with quantum processors. In more complex workflows, quantum subtasks can form a portion of a larger total simulation that to a large extent utilizes classical computing resources. Thus, the balance between classical and QC varies widely depending on the problem at hand.

The design and implementation of integrated traditional HPC and QC systems emerged only recently. 
Integrating quantum computers into existing HPC infrastructures presents several technical challenges~\cite{Britt2017,Johansson2021,humble2021quantum,Elsharkawy2023,elsharkawy2023integrationQuantumAccelerators,elsharkawy2024integration,schulz2023accelerating}. First, QC is a nascent and therefore still limited resource, currently accessible to a relatively small user base, mostly via the cloud. Second, present QC paradigms are largely incompatible with traditional HPC programming standards. Third, contemporary quantum computers lack the power and scale needed for most practical applications (see Sec.~\ref{use_cases}). Consequently, the computational demands placed on classical resources in hybrid systems are still minimal compared to what supercomputers can provide. Fourth, the broad spectrum of quantum hardware architectures (see Sec.~\ref{platforms}) currently necessitates specialized control software for each implementation. And fifth, the rapid development of quantum hardware further complicates integration, as investments in hybrid systems have a high risk of quickly becoming obsolete.

Despite the challenges outlined above, numerous HPC centers are actively exploring the integration of HPC and QC, even as current quantum computers have yet to achieve the performance benchmarks set by HPC standards. This is often driven by recognizing the potential use of quantum-enhanced HPC for the user base in the future. The following paragraphs highlight the innovative efforts undertaken at various facilities affiliated with ADAC members, showcasing their commitment to advancing this evolving field.\footnote{The list is not exhaustive}.

The EuroHPC Joint Undertaking has launched initiatives for integrating several different QC modalities into existing supercomputing infrastructures. The first of the initiatives, HPCQS, integrates two neutral atom simulators to European supercomputers~\cite{EuroHPC-HPCQS-2022}. Subsequently, eight other quantum computers have been announced throughout Europe~\cite{EuroHPC-6QC-2023, EuroHPC-2QC-2024}. The first installations \changes{were finished in 2025, with broad accessibility expected during 2026. The EuroHPC Federation Platform (EFP), a secure one-stop portal that federates identity and core services so users can seamlessly access and manage resources across Europe’s heterogeneous EuroHPC supercomputers, is set to incorporate the EuroHPC quantum systems as well~\cite{EFP}.}

The Finnish Quantum-Computing Infrastructure (FiQCI) has integrated the EuroHPC LUMI supercomputer, hosted by CSC -- IT Center for Science, with several quantum computers~\cite{FiQCI}. Access to the first HPC+QC service was opened to users in 2022, and in 2025, a complete hybrid setup integrating full HPC capacity with the national fifty-qubit quantum computer, hosted by the Technical Research Centre of Finland (VTT), was launched~\cite{LUMI-Q50}. The mission of FiQCI is to provide state-of-the-art quantum-computing services such as computing time and training to Research, Development, and Innovation (RDI) communities. This includes providing a hybrid HPC, artificial intelligence (AI), and QC (HPC+AI+QC) platform for developing, testing, and exploiting quantum-accelerated computational workflows. An experimental AI-optimized QC platform, LUMI-IQ, is included in the upcoming EuroHPC LUMI AI Factory, hosted by CSC -- IT Center for Science~\cite{LUMI-AI-Factory}. LUMI-IQ aims to tightly integrate classical compute capacity, AI tools, and QC in one unified, highly performant HPC+AI+QC platform.

Drawing on its HPC expertise and HPC lifecycle management of the US Department of energy (DOE), as well as the Quantum Computing User Program (QCUP)~\cite{QCUP}, Oak Ridge National Laboratory (ORNL) proposes a hardware-agnostic framework for augmenting classical HPC with QC capabilities~\cite{beck2024integrating,shehata2025bridging}. 
\changes{Recently, ORNL has installed several different types of QC technology on-premises to test and evaluate these integration paradigms. The first installation was a triplet of quantum computers developed by Quantum Brilliance, which uses nitrogen-vacancy centers in diamond to realize room-temperature operation. These systems have been installed locally at ORNL alongside conventional HPC infrastructure. As part of the lab’s Heterogeneous Quantum Systems Initiative~\cite{HQS_Initiative}, researchers are using the Quantum Brilliance’s systems to evaluate tightly integrated hybrid quantum-classical workflows~\cite{QB_ORNL}. These efforts have established additional software frameworks to deploy and manage distributed HPC+QC applications, with an initial application focus on quantum chemistry calculations that take advantage of classical parallelism. The second installation is a superconducting quantum computer delivered by IQM~\cite{IQM_ORNL}. The 20-qubit transmon system provides a unique opportunity to operate cryogenic computing systems alongside conventional HPC infrastructure. Researchers are integrating the IQM system with similar hybrid quantum-classical workflows as a path to strengthening convergence between quantum computing and HPC technologies. A key use case is the investigation of quantum error correction methods, especially syndrome decoding, using both quantum and classical control techniques. ORNL has integrated a Riverlane DeltaFLow decoder within the computing enclave~\cite{Deltaflow}. The DeltaFlow decoder is a dedicated computational system for decoding syndromes for QEC methods, and the ORNL efforts are evaluating the performance of the complete HPC+QC ecosystem. This is echoed by recent collaborations with industry on integrated architectures, including NVIDIA’s NVQLink\footnote{co-developed by NVIDIA in partnership with ORNL and three other DOE labs, Pacific Northwest National Laboratory (PNNL), Lawrence Berkeley National Laboratory (LBNL) and Sandia National Laboratories (SNL), all of which are ADAC members or affiliates~\cite{ADAC}} and CUDA-Q tools~\cite{NVIDIA_ORNL,NVQLink}, which support the ORNL goal to harmonize diverse quantum technologies with leadership-class supercomputing. These integration efforts are further amplified by the on-going development of software methods and architectures that can accommodate hybrid workflows. These goals are central to the Quantum Science Center, a National Quantum Information Science Research Center established by the U.S. Department of Energy to accelerate advances in quantum computing by leveraging leadership-class HPC~\cite{QSC}. The multi-institutional collaboration coordinates efforts between the DOE national laboratories, universities, and industry to realize scalable quantum-centric HPC systems and applications. The center works closely with the openQSE, open quantum software ecosystem, working group to develop best practices for hybrid computing systems to address these scientific goals~\cite{OpenQSE}.}

Still within DOE, the Pacific \changes{Northwest} National Laboratory (PNNL) has recently launched an initiative on hybrid HPC+QC computing, focusing on algorithm and application development for scientific computational domains such as quantum chemistry, climate transportation, and electric grid optimization. 

In Japan, AIST has established the Global Research and Development Center for Business by QC-AI technology (G-QuAT) in 2023 to build a global quantum industry ecosystem, including use case creation, supply chain resilience, and workforce development. ABCI-Q is their hybrid AI+QC computing infrastructure that integrates an AI supercomputer and three types of on-premise QPUs: Fujitsu’s superconducting, QuEra’s neutral atom, and OptQC’s photonic. Still in Japan, RIKEN, Softbank Corp., the University of Tokyo, and Osaka University are carrying out the JHPC-Quantum project~\cite{JHPC-Quantum} supported by NEDO, a Japanese agency under the Ministry of Economy, Trade and Industry. In this project, a platform that connects multi-site supercomputers and different types of quantum computers is developed, aiming at early commercialization of quantum computers.

The Australian Supercomputing Innovation Hub located inside the Pawsey Supercomputing Centre has been experimenting with integration of on-prem, off-prem, and emulators with the HPC systems. Completed projects include integration of a room temperature diamond-based quantum computer with a Cray EX system, a mix of simulators and emulators running on AMD and NVIDIA GPU nodes exposed as virtual QPUs, and the ability to launch remote jobs on AWS Braket and a neutral atom QuEra system.

In India, the Center for Development of Advanced Computing (C-DAC) is advancing HPC+QC integration by leveraging PARAM supercomputers with GPU, FPGA, and Vector accelerators to enhance QC simulations and hybrid workflows. Qniverse~\cite{Qniverse} developed by C-DAC provides an integrated platform for designing, simulating, and running quantum circuits, offering a visual interface for circuit creation, hardware resources like CPUs, GPUs, and QPUs, along with tools for job management, gate operations, and quantum algorithm exploration.

\changes{Across these efforts, the main differentiators lie not in the existence of hybrid workflows but in how they are operationalized: the location of the QPU (on-premises vs. remote), the latency and co-scheduling assumptions underlying the workflow, and the responsibility for managing the control stack and telemetry. For HPC centers, these choices directly shape integration efforts related to identity/access, queuing/accounting, software environment management, and reproducibility.}

\changes{HPC+QC integration can be described along two complementary axes. The first is the \emph{integration layer}: (i) \emph{hardware integration}, which concerns where the QPU is deployed and how it is connected to classical resources; (ii) \emph{software integration}, which covers programming and execution models, compilation, orchestration, and scheduling across classical and quantum components; and (iii) \emph{operational integration}, which includes access control, accounting, monitoring, and reproducibility in production environments. 
\\
\indent The second axis is the \emph{deployment model}: in \emph{loose integration}, the QC resource is accessed remotely (typically via cloud services and APIs), whereas in \emph{tight integration}, the QPU is co-located with the HPC system and can, in principle, support lower-latency coupling and stronger co-scheduling. In practice, most present-day workflows primarily require software and operational integration, while tight coupling becomes increasingly important for latency-sensitive feedback loops (e.g., real-time QEC). 
\\
\indent In the following, we first discuss hardware and software integration, and later return to the comparison of loose versus tight integration from an operational and benchmarking perspective.}

\subsection{Hardware Integration}
\label{HW integration}
The hardware level of integration involves the deployment of physical resources through two primary approaches: loose integration with remote connection between HPC and QC, as well as tight integration with co-location of both compute systems~\cite{Johansson2021,Valeria2021QuantumForHPC,beck2024integrating,shehata2025bridging}. 

\begin{figure}[h]
 \includegraphics[width=0.4\textwidth]{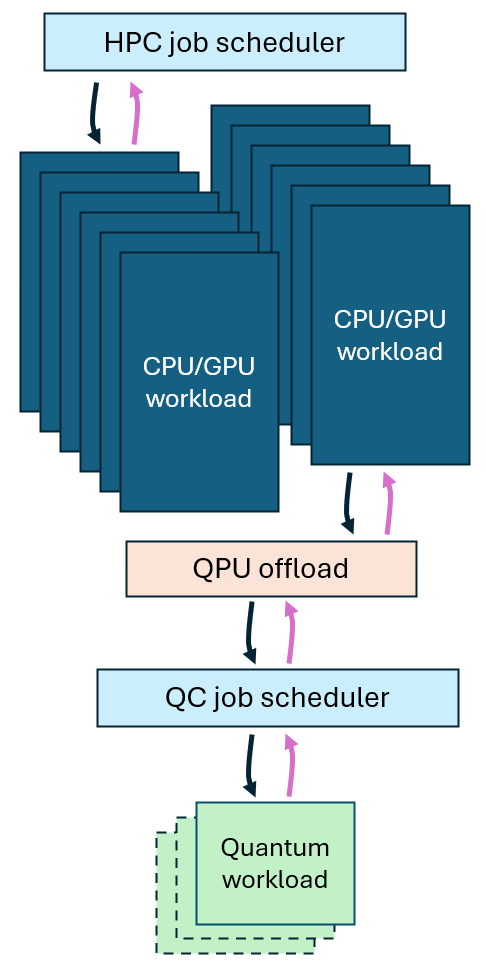}
 \caption{\label{fig:HPC+QC} 
 Generic execution flow of a hybrid classical HPC+QC task.
}
\end{figure}

\subsubsection{Loose Integration Model}
\label{loose integration}
In a loose integration model, classical and quantum hardware are located at different premises and connected through a network interface or via the cloud~\cite{Johansson2021,Valeria2021QuantumForHPC}. The benefit of this approach comes from increased modularity and flexibility. Loose integration allows more straight-forward connection of new quantum resources to the HPC infrastructure while minimizing disruptions to the HPC users. On the other hand, the distance between physical resources introduces additional latency of communication between the HPC and QC installations. Presently, communication latency is of negligible consequence, as other processes such as quantum hardware initialization are much more time consuming than data transfer between HPC and QC. Promising NISQ-era algorithms which rely on iterating information between classical and quantum resources could potentially be limited by latency in the future, however. Having components of the infrastructure at different premises or providers does add a layer of complexity when it comes to deploying components such as schedulers or resource management systems. Most of the potential drawbacks of the loose integration model can be addressed by deploying intermediate classical resources on-site with the quantum computer~\cite{Johansson2021}.

\subsubsection{Tight Integration Model}
\label{tight integration}
Tight integration -- seamless coupling of quantum and classical computing resources within a unified system architecture -- has its advantages. However, a flurry of technical challenges need to be addressed to enable the two paradigms to work in harmony with minimal overhead and maximal performance. \changes{One of the primary challenges is the lack of a standardized interface that couples classical HPC systems to a QPU within the timing budget of latency-sensitive feedback loops (e.g., real-time QEC, calibration, or adaptive circuits). While efforts such as NVIDIA NVQLink~\cite{NVQLink,caldwell2025platformarchitecturetightcoupling} aim to standardize GPU-QPU integration (usually in the context of QEC with syndrome extraction, decoding and correction), microsecond-scale interconnect latencies are challenging for fast-cycle platforms like superconducting and photonic qubits, whereas slower technologies such as neutral atoms or trapped ions are less constrained.} 
\\
\indent Among other hurdles, current quantum computers can have large physical footprints and require frequent calibration and maintenance, often necessitating on-site personnel. Also, many QPU platforms operate under stringent environmental conditions, and can require cryogenic systems such as dilution refrigerators and shielding from electromagnetic noise. Additionally, the rapid pace of advances in quantum hardware renders quantum hardware obsolete on a much faster timescale than traditional processor technology. This necessitates regular upgrades to the QC infrastructure in order to remain competitive.

\subsection{Energy Efficiency}
\label{energy efficiency}
The hardware level of integration also needs to be considered from the perspective of \changes{full-system} energy efficiency. Indeed, while quantum computers hold immense promise for industry and society, their \changes{\emph{end-to-end} energy footprint is still not consistently characterized} in current deployment strategies~\cite{efficiency_Auffeves_2022,fellous2023optimizing,Efficiency_Nature_2023}. \changes{For hybrid HPC+QC workflows such as presented generically in Fig.~\ref{fig:HPC+QC}, energy consumption is not determined by the QPU alone: it can include the QPU environment (e.g., cryogenics or vacuum systems), control electronics and readout hardware, classical low-latency compute used for calibration and decoding, and facility overheads (power distribution, cooling, and infrastructure).
}

To address this challenge, research and industry experts across diverse fields—including quantum physics, technology, hardware and software—have recently started to unite to better understand the physical resource demands of emerging quantum technologies~\cite{ikonen2017energy, smierzchalski2024efficiency,jaschke2023quantum,martin2022energy,meier2023energy, aamir2025thermally}. This has led to initiatives like the Quantum Energy Initiative~\cite{QEI} as one example, with more likely to emerge as the field grows. 

\changes{Quantum computers may in principle provide a “green quantum advantage” by achieving a lower \emph{energy-to-solution} than classical hardware for a given computational objective. However, whether such an advantage materializes is inherently conditional: it depends on the full workflow, the required number of circuit repetitions (sampling), compilation and error-mitigation overheads, and the amount of classical pre-/post-processing in the hybrid loop. A holistic view that accounts for the \emph{total} energy consumption of the combined HPC+QC workflow is therefore required to determine whether QC can deliver an energy advantage in addition to any computational benefits for real-world applications.}

\changes{From an HPC-center perspective, this motivates the use of operationally meaningful metrics, e.g., energy-to-solution for representative workloads, utilization-aware energy accounting (including idle/standby power), and facility KPIs such as Power Usage Efficiency (PUE)~\cite{ISO-PUE:2026} and, where relevant, Water Usage Efficiency (WUE)~\cite{ISO-WUE:2022}. Establishing shared measurement boundaries and reporting practices is essential for meaningful comparisons across integration models and sites.} 

\changes{These considerations naturally extend benchmarking from device-centric metrics toward workflow-centric measures that include energy, overheads, and operational efficiency (see Sec.~\ref{benchmarking}).}

\subsection{Software Integration}
\label{software integration}
The software level integration in QC is a complex process that involves multiple components. In quantum-centric supercomputing\changes{~\cite{alexeev2023quantumcentric}}, the software stack should encompass application-level quantum algorithms, scheduling of both quantum and classical subroutines, compiling of quantum programs, and hardware-level control of quantum computer. Optimized software stack is important to reduce the runtime and improve the accuracy of hybrid HPC+QC \changes{workflows (see Fig.~\ref{fig:HPC+QC}}). For example, efficient compilers and routing algorithms can help to reduce runtime for transpilation, as well as create more optimized circuits~\cite{elsharkawy2024integration} tailored for specific quantum computers. On the other hand, post-processing methods like error mitigation are required to deal with errors caused by noisy quantum hardware. Efficient control and scheduling layers are crucial to avoid bottlenecks and maximize QPU utilization, ensuring balanced and optimal use of quantum and classical resources (Sec. \ref{ssec:orchestration} and \ref{ssec:scheduling}). This multilevel integration is challenging from the perspective of standardization.

\subsubsection{Software Stack}
\label{software stack}

Research and development in software is a key aspect in the maturing of QC. Software is indispensable in virtually all areas related to QC, from quantum control (the operation of the computers themselves) to quantum algorithms, numerical simulations, and hardware modeling. Quantum computers currently present themselves predominantly as accelerators embedded in a mostly classical computational stack. This implies programming and execution models that lead to an increase in heterogeneity resulting from the introduction of quantum computers. In this context, robust software is paramount in offsetting the costs in introducing new abstraction layers and compilation approaches necessary to fully reap the benefits of turning to quantum computers.

The fundamental differences between conventional computers as embodied by HPC facilities and quantum computers lead to a myriad of challenges when considering the integration of these two technologies. These concerns range from the physical integration itself (e.g. network interconnects) to the proper software stack that is able to manifest the full potential of this synergy. Against this backdrop, we highlight three main aspects that underlie the quantum software stack concerns from the HPC standpoint, namely, execution and programming models, orchestration, and scheduling. These aspects ultimately shape the overall software stack and how it interacts with external software, such as libraries and software development kits (SDK).

\subsubsection{Execution Models}
\label{ssec:execution_models}

We can imagine the HPC system interacting with one or several quantum computers in one of three basic models: 1) client-server; 2) quantum-accelerated nodes via classical network; 3) quantum-accelerated nodes via classical and quantum network. These models follow an order of increasing complexity, but also potentially higher computational power.  We further elaborate on these models in the following paragraphs. Interested readers are referred to Refs.~\cite{Britt2017, elsharkawy2023integrationQuantumAccelerators, Elsharkawy2023, elsharkawy2024integration} \changes{for a more detailed discussion}.

\changes{The client-server model (1) facilitates interaction with quantum computers by enabling access through standard HTTP protocols and vendor-provided RESTful APIs.} The client system carries out the processing necessary to generate quantum circuits, which are transmitted over the cloud to the server that hosts the quantum computer. \changes{The server executes the circuits and provides the results for the client, which is responsible for processing the output.} This approach offers the lowest level of integration, as the two systems are spatially separated, which implies little to no modification to the corresponding software stacks. However, because the two compute systems are also, in some sense, temporally separated, this integration modality suffers from comparatively high latency due to the long-distance communication. This is exacerbated 
\changes{by a double-queue effect (waiting in both the HPC batch system and the provider-side quantum queue)}, as execution in both client and server may depend on a queuing system.

In (2), one can envision a compute node whose new component is a QPU, akin to GPUs that are now ubiquitous in current HPC nodes. However, inter-node communication is limited to classical communication over typical HPC network, or, in other words, there is no explicit communication across QPUs. One of the main benefits of this scheme is that it can drastically reduce the latency in CPU-QPU communication, as proposed by the NVIDIA DGX Quantum \changes{platform}~\cite{dgx}. This model presents a larger boundary across the two software stacks, leading to changes in scheduling, and more importantly, in the overall orchestration. However, 
\changes{without inter-QPU quantum communication, scaling is primarily in throughput (running more independent quantum tasks in parallel) rather than enabling a single distributed quantum state across nodes; thus speedups for a single tightly-coupled problem are limited.}~\cite{Britt2017}.

Finally, (3) augments (2) by enabling inter-node quantum communication via quantum networking, \changes{so that quantum states can be distributed across multiple QPUs. This model is valuable because it can increase the effective problem size that can be represented coherently, enabling distributed quantum algorithms and modular architectures~\cite{barral2024_DQC}. Speedup, however, is not automatic: it is algorithm-dependent and can be limited, or even negated, by the overheads of entanglement generation and distribution, error rates, synchronization and control complexity, and the cost of moving classical data between nodes. As a result, this deployment model is the most demanding in terms of hardware, software, and operational infrastructure, and its practical benefits will hinge on tight co-design of networking protocols, fault tolerance, and workflow orchestration.}

\subsubsection{Programming Models}
\label{programming models}
These basic elements of the execution models (Sec. \ref{ssec:execution_models}) define the structure of the programming model. 
\changes{It is anticipated that most programming model innovations will arise at the interface between HPC and QC, as programming approaches for each stack independently are expected to remain largely unchanged.}

Programming in such a hybrid fashion would entail putting forth abstractions that can adequately handle the intricacies of each compute paradigm, allowing for interoperability across the myriad of emerging QPU technologies and providers. Parallel computing being one of the key characteristics that elevated HPC to its current status, proper communication across the different components of the HPC+QC stack will be of paramount importance
\changes{. This }is exemplified by efforts in extending the message passing interface (MPI) standard to enable quantum communication~\cite{haner2021distributed}.

In light of the discussions involving Orchestration (Sec. \ref{ssec:orchestration}) and Scheduling (
\changes{Sec}. \ref{ssec:scheduling}), programming models need to be developed to enable robust asynchronous task execution across the 
\changes{HPC+QC} boundary. In some instances, circuit cutting and subsequent knitting may be 
\changes{advantageous or required to manage limited resources. This necessitates careful consideration within the programming model and must be incorporated into the underlying orchestration and scheduling strategies.}

\subsubsection{Orchestration}
\label{ssec:orchestration}

\changes{In the current client-server model, HPC resources are rarely needed or considered secondary, with orchestration primarily managed by the end user. The user determines the division of workflow tasks between the client and server based on the target algorithm. Due to minimal demand for complex orchestration, there is little requirement for specialized tools for workflow composition and management.} As the two stacks work almost independently of one another, orchestration across the 
\changes{HPC+QC} boundary is almost entirely established at compile-time.

As HPC takes a more prominent role in workflows involving QPUs, there will be an increasing need to adopt orchestration best practices. For example, automated workflow analysis to map tasks onto classical and quantum “primitives” without direct intervention from the end user is an immediate orchestration concern. 
\changes{Furthermore, by revealing physical details from the hardware, an orchestrator in this context would be capable of determining which of the available QPUs to offload a specific task to.}

\subsubsection{Scheduling}
\label{ssec:scheduling}

\changes{The scheduler's role is to optimize QPU utilization, or equivalently minimize idle time. However, leveraging flexible, backend-agnostic software stacks poses challenges due to significant variability in QPU technologies, including differences in operational timescales, circuit synthesis, and calibration frequency and duration.}The most obvious direction seems to be extending current industry standards, such as SLURM and PBS, with the functionality necessary to manage scheduling across QPUs and classical HPC.

\subsubsection{Software Development Kits}
\label{SDK}

Quantum \changes{SDKs} provide essential tools for developing and simulating quantum algorithms, integrating classical and QC paradigms. 
\changes{The development objectives of the software libraries span from validating quantum hardware through error models at either an abstract or device-specific level, to simulating hardware-agnostic high-level quantum logic operations that strive to efficiently approach the boundary of quantum advantage.} Many libraries offer frameworks for quantum circuit design, optimization, and execution on QPUs or simulators, supporting various quantum programming models such as gate-based and variational approaches, facilitating research in quantum ML (QML), cryptography, and optimization. Hybrid computing frameworks like XACC~\cite{McCaskey2018XACC} enable seamless execution across classical HPC and quantum processors. QC simulation libraries offer a portfolio of different simulator types, such as full statevector and tensor network methods \changes{(see Sec.~\ref{ssec:full_state_methods)})}. As QC progresses, open-source quantum libraries play a crucial role in standardizing quantum software development and accelerating advancements in the field~\cite{Preskill2018quantumcomputingin}. 

\changes{Some of the most commonly used QC libraries are listed below.} Note that many open-source quantum SDKs have lot of similarities in terms of supported features, and listing them all would be impractical. For example Qiskit (IBM)~\cite{Qiskit} and Cirq (Google)~\cite{Cirq} both provide their own set of tools for different parts of QC workflows, such as circuit creation, transpilation and circuit execution. 
\changes{For a more comprehensive list of quantum software libraries, the reader is referred to }Ref.~\cite{quantumsimslistgithub}.

\subsubsection*{Qiskit}
Qiskit~\cite{Qiskit} is a software stack for QC \changes{developed by IBM}, providing tools for quantum circuits, operators and primitives, as well as circuit transpilation methods, quantum workflows and simulators with realistic or custom noise models. Qiskit and its add-ons provide application level frameworks for applications such as natural sciences, finance and ML.

\subsubsection*{Pennylane}
Pennylane~\cite{bergholm2022pennylaneautomaticdifferentiationhybrid}\changes{,} developed by Xanadu, is a cross-platform Python library for differentiable programming of quantum computers. This type of quantum programming is well suited for trainable quantum circuits, with applications in QML and quantum chemistry.

\subsubsection*{CUDA-Q}
NVIDIA CUDA-Q~\cite{cuda-q-url} is 
\changes{a} QPU-agnostic platform for accelerated quantum supercomputing.
CUDA-Q provides tools for hybrid quantum classical computing, \changes{a low-level virtual machine (LLVM)}-based quantum compiler, and multiple simulators optimized for NVIDIA's GPU systems.

\subsubsection*{TKET}
TKET and TKET2~\cite{tket-url}
\changes{are} software platforms \changes{designed} for the development and execution of gate-level quantum computation, providing state-of-the-art performance in circuit compilation.
TKET involves \changes{a} compiler 
\changes{that includes} qubit mapping and routing, several optimization paths, quantum circuits executions on different types of QPUs, and simulators.   
TKET2 is an open source quantum compiler.

\subsubsection*{Intel Quantum SDK}
\changes{The Intel Quantum SDK~\cite{intel-qsdk-url} provides a comprehensive quantum computing stack in simulation, designed to eventually support Intel’s spin-qubit chips. It includes components such as a C++ quantum compiler, quantum runtime, generic simulators, and Intel-specific hardware simulators. The compilers are built upon the LLVM framework.}

\subsubsection*{Qualtran}
Qualtran (quantum algorithms translator) is an open-source library for representing and analyzing quantum algorithms developed by Google~\cite{harrigan2024expressing}. It incorporates a set of abstractions for representing quantum programs and a library of quantum algorithms expressed in that language to support quantum algorithms research. Qualtran has an internal integration of resources estimation~\cite{qualtran}.


\subsection{Artificial Intelligence for Quantum Computing}
\label{AI for QC}
AI and QC reinforce each other in a symbiotic loop. QC can accelerate and enhance traditional AI processes, while AI is increasingly needed for the efficient operation and utilization of quantum computers. AI will play a pivotal role in the full hybrid HPC+QC software stack. There is an increasing demand for classical pre- and post-processing for creating the input for the quantum computer to process, and for analyzing and refining the data coming out of a quantum computer. The complexity of these classical tasks increases rapidly with increasing qubit count. Thus, the more powerful a quantum computer is, the more classical computing resources are required to run it efficiently. It is expected that the classical tasks can be significantly sped up by utilizing ML and AI.

HPC has a much broader role than just taking care of the non-quantum-accelerable parts of AI, however. HPC is needed for several pre- and post-processing tasks during different stages of the entire computational process. Some of these are:
\begin{itemize}
    \item Optimal compilation, transpilation, circuit knitting, and qubit routing of quantum circuits
    \item Design and creation of quantum algorithms
    \item Error mitigation and enhancing signal-to-noise ratio of QC output
\end{itemize}

All of the above  examples are prime candidates for ML and AI exploitation. Using AI techniques, the overall computational procedure can thus be significantly enhanced.

It might be instructive to think of a quantum computer as a measurement device for quantum experiments. In the case of computing, the experiment happens to be a sequence of logical operations on quantum states. To get the quantum computer to perform a set of instructions, the algorithm description, that is, the higher-level quantum programming language needs to be compiled to the instruction set of the quantum computer. There are a few points that differentiate compilation for a quantum computer from classical compilation.

When compiling a quantum program, one has to account for efficiency in a significantly stricter manner. Due to decoherence and inherent errors, the longer the compiled or transpiled quantum circuit is, the less reliable the computed result becomes. In a classical computer, an inefficient compilation or an inefficient solution to a given problem will just take longer to complete, but the result should essentially be identical. 
\changes{For a quantum computer, an inefficient quantum circuit may lead to unreliable or incorrect outputs, often referred to as "garbage", due to cumulative errors, decoherence, and the lack of proper optimization in the circuit design.} It is therefore important to efficiently optimize the compilation process. This can become a very costly optimization problem. Ideally one needs to consider that different qubits have different coherence times, operations on different pairs of qubits have different error rates, and operations involving qubits that are not directly connected to each other require additional operations that all add to the execution time and total error of the calculation. In a classical computer, compilers can rely on all transistors working with equal reliability, in a quantum computer, all qubits in practice are better or worse.

Designing quantum algorithms is naturally a highly demanding task in itself, as evident by the still rather limited set of primitive quantum algorithms \changes{currently available}. Combining quantum primitives to solve some given higher-level problem is almost equally challenging as coming up with a new algorithm joining the likes of Shor’s, Deutsch-Jozsa, Harrow–Hassidim–Lloyd (HHL), and quantum phase estimation algorithms. Recent developments in utilizing Large Language Models (LLMs) for designing quantum algorithms have the potential to revolutionize quantum programming, again in a more fundamental way than how AI and LLMs are transforming classical programming~\cite{Liang2023,nakaji2024generative,Rudolph2023}. In algorithms design, the use of generative pre-trained transformers such as GPTs in the design of new quantum algorithms is emerging, including the Generative Quantum Eigensolver~(GQE)~\cite{nakaji2024generative,minami2025gqco}. GQE utilizes a transformer architecture to generate sequences of \changes{a} quantum circuit instead of a parameterized quantum circuit such as VQE.

A significant source of error stems from the readout process, which refers to the extraction of results from a quantum computation. Minimizing readout errors is therefore crucial to enhancing the usability of both Noisy Intermediate-Scale Quantum (NISQ) computers and Fault-Tolerant Quantum Computers (FTQCs).

ML and AI tools can also be used to improve the currently noisy outputs of quantum computers. 
\changes{A significant source of error stems from the readout process, which refers to the extraction of results from a quantum computation.} Therefore, it is essential to minimize the readout-error in order to reach the usability of both NISQ computers and FTQCs. In this respect ML approaches have been proven effective in characterizing the response functions of the detectors of quantum computers based on superconducting qubits. Error mitigation~\cite{liao2023machine} and correction~\cite{zeng2023approximate,CSIRO_AI_QEC_2024,bausch2024learning,acharya2024quantum} procedures based on ML and AI are emerging as crucial components of an efficient QC software stack~\cite{Torlai2023,Filippov2023}. \changes{In this context, the tight integration model discussed in Sec.~\ref{tight integration}, and in more details in Sec.~\ref{real_time} and \ref{maturity int mods}, is particularly relevant: dedicated HPC resources can be coupled closely with the quantum machine, effectively forming part of the vendor-provided system, to provide sufficient classical compute for running AI-enhanced decoding, compilation, and other hybrid workflows that require low-latency interaction between the classical and quantum components.}

In the 
\changes{NISQ} era of QC, the main factor limiting practical applications is the number of quality qubits available on a single QPU. 
\changes{To realize quantum advantage on NISQ and future devices, large quantum algorithms can be divided into smaller parts for execution either in parallel on multiple QPUs or sequentially on a single QPU. This technique, known as circuit cutting or circuit knitting, helps overcome hardware limitations, particularly the limited number of qubits in NISQ devices, and reduces the errors caused by circuit depth and decoherence. However, it increases the sampling overhead and can amplify noise effects. Finding the optimal way to split a circuit becomes increasingly challenging as the number of qubits grows.} Finding the optimal \changes{balance for dividing} a given quantum circuit (algorithm) becomes a highly complex problem, especially with increasing qubit count. AI tools for optimal circuit knitting and cutting will be needed in order to make this approach feasible on larger quantum devices.

AI can advance the field of QIS in several other areas~\cite{alexeev2024AIQC}. For example, AI can significantly improve the scalability of quantum state estimation and dynamics approximation by reducing the extensive memory demands of large quantum systems. Neural Quantum States (NQS) approximate quantum wave functions using neural networks~\cite{gao2017efficient, carrasquilla2021manybody}, and can be used for simulations of quantum many-body dynamics and predictive modeling of observables without explicitly storing quantum states~\cite{krenn2023artificial}. In quantum metrology, AI can improve and advance parameter estimation and measurement strategies (both non-adaptive and adaptive)~\cite{krenn2023artificial}. AI's contribution to adaptive measurement strategies is particularly notable, where neural networks and other ML techniques enhance the accuracy of quantum simulations~\cite{quek2021adaptive}.

These studies collectively demonstrate the transformative potential of AI in advancing QC technologies, paving the way for practical and scalable hybrid computation that combines HPC, AI, and QC.


\subsection{Real-time QEC and HPC}
\label{real_time}
Moving beyond today’s NISQ devices toward early FTQC requires QEC, as explained in Sec.~\ref{FTQC}. In QEC, quantum information is encoded across many physical qubits to form \emph{logical} qubits, and the system repeatedly measures parity checks to extract \emph{syndromes}: bits that indicate which checks ``fired'', revealing likely error locations without directly measuring (and thus collapsing) the quantum state. A classical decoder processes the syndrome stream and outputs either explicit corrections, or a \emph{Pauli-frame} update, thereby suppressing noise while computation proceeds.

In this context, \emph{real-time} primarily means sustaining the \emph{syndrome-processing throughput} of repeated QEC cycles and keeping the measurement-to-decode-to-action path within the timing budget set by the hardware and the chosen code family. 
Qubit modality strongly shapes these requirements. Fast platforms with short cycle times (e.g., superconducting and photonic approaches) can demand microsecond-scale, low-jitter decoding and control paths, pushing the decoding compute to behave like instrumentation rather than something amenable to scheduling. Slower platforms (e.g., trapped ions and many neutral-atom implementations, often with two-qubit gates in the $\sim$10--100~$\mu$s range) can relax hard real-time constraints and shift attention toward orchestration, calibration drift, and data movement overheads; timing budgets may tighten as control techniques mature (see Table~\ref{fig:hardware_landscape}). Thus, the required throughput is platform-dependent and can change across hardware generations.

Two representative implementation directions illustrate how close-to-QPU classical compute may be provisioned. i) Nvidia’s NVQLink concept \cite{NVQLink,caldwell2025platformarchitecturetightcoupling} emphasizes integrating QPUs and GPUs through a low-latency, high-throughput link for efficient quantum-classical communication. The host GPU-node is dedicated, with features like pinned resources, low jitter, controlled updates, and contention-free networking, resembling an instrument control node rather than a general-purpose cluster. ii) IBM has highlighted FPGA-based real-time decoding, where deterministic latency and tight hardware control can simplify, meeting microsecond deadlines, albeit typically at the cost of more specialized toolflows and deeper co-design across the QPU control stack and facility operations~\cite{IBM-FPGA-decoding-blog, muller2025improvedbeliefpropagationsufficient, maurer2025realtimedecodinggrosscode}. Similarly, Riverlane has reported strong results using FPGAs for decoding~\cite{Ziad2025}, demonstrating their potential in optimizing error correction performance.

For slower qubit architectures, timing requirements for the decoding process are relaxed. IonQ has shown how even a standard CPU without any parallelization could be sufficient for current trapped ion and neutral atom systems, and that less than a hundred CPU-cores would be able to handle decoding for a 1000 logical qubit system~\cite{ye2025beamsearchdecoderquantum}.

Current demonstrations are, however, far below the scale required for high impact, fault-tolerant applications. While hardware decoders like FPGA may benefit current conceptual demonstrations, the scalability of the hardware requirements is highly unfavorable for large-scale system development due to the number of control lines as well as space and power usage. More advanced microelectronic solutions are expected to be necessary for scalable and resilient QC systems.

From an HPC-center perspective, it is useful to separate what must run \emph{in the hard real-time loop} from what can run on general HPC resources. A practical tiering is:
\begin{itemize} 
    \item \textbf{In-loop, hard real-time decoding (edge tier):} This tier must keep up with syndrome generation continuously. In practice it requires dedicated classical compute near the QPU control stack (often the same rack row or system enclosure), using FPGAs/ASICs and/or GPU “real-time hosts” with kernel-bypass networking and carefully managed jitter. NVQLink-style architectures~\cite{caldwell2025platformarchitecturetightcoupling} explicitly target this regime by making the accelerated node part of the logical QPU system. 
    \item \textbf{Near-real-time services (seconds to minutes):} Many tasks that strongly affect QEC success do not need microsecond closure: decoder parameter refresh, calibration analysis, noise-drift detection, layout/compilation updates, and online performance monitoring. These can run on a small reserved HPC partition (CPU/GPU) without strict determinism, feeding updated models and parameters back to the edge tier. 
    \item \textbf{HPC at scale (hours to days):} Large-scale simulation, decoder benchmarking, training/validation of AI-assisted decoders, sensitivity studies, and log analytics are natural HPC workloads. The practical value is high because real-time QEC performance is often limited by rare events, correlated noise, or miscalibration. These phenomena benefit from large ensembles and extensive analysis rather than only faster edge hardware. \end{itemize}

Only the first tier is truly latency-critical; the latter two are typically where established HPC workflows and scale provide the highest leverage.

Scaling further increases the classical burden in ways familiar to HPC practitioners. Every QEC round produces streaming data that must be ingested, routed, decoded, and turned into updates continuously. As physical qubit counts rise, aggregate data rates and bookkeeping can become first-order design constraints. Increasing code distance (the number of physical qubits involved in implementing a fault-tolerant logical qubit to detect and correct multiple errors~\cite{Fowler_2012}) to improve protection also grows the decoding problem size and increases the number of physical qubits per logical qubit. For surface-code layouts discussed in recent experiments, scaling is often described as on the order of $2d^2$ physical qubits per logical qubit. In practice, the classical system evolves from a single fast decoder into a parallel service that must scale horizontally (multiple decoders/accelerators, more bandwidth, and more buffering) while preserving determinism where required for control.

Finally, AI is increasingly explored for QEC decoding and noise adaptation: learned models can map syndrome streams to likely errors while implicitly capturing device-specific effects (e.g., leakage and correlated noise) that are difficult to model explicitly. Also off-line QEC circuit compilation including placement and routing is amenable to AI-enhancement. This typically splits into an HPC-friendly workflow: training/validation and large-scale evaluation on HPC resources, while low-latency inference is deployed at the edge near the QPU to meet real-time throughput goals. Overall, realistic QEC deployments at scale tend to converge on a tiered classical architecture: dedicated, deterministic edge compute for in-loop decoding, complemented by conventional HPC resources that improve QEC indirectly through monitoring, tuning, and large-scale analysis.

\subsection{Comparison of integration models}
\label{maturity int mods}
Hybrid quantum--classical workflows are typically realized through two integration models as discussed above: (i) \emph{tight integration}, where a quantum computer or QPU is co-located with the HPC system and connected via a local high-performance network; and (ii) \emph{loose integration}, where the QC is accessed remotely through internet-facing services and APIs. Both support the same basic pattern, classical pre/post-processing on HPC combined with quantum circuit execution, but they differ in software control, operational responsibilities, and the feasibility of latency-sensitive feedback loops.
In the \emph{loose} model (Section~\ref{loose integration}), the QPU is usually a provider-managed service. This reduces on-site deployment and maintenance effort and can speed access to new features as platforms evolve. The trade-off is a split software stack across administrative domains: the HPC center controls the classical environment, while the provider controls parts of the quantum execution environment. This can limit the ability to enforce uniform runtime assumptions and complicate reproducibility, which then relies on stable interfaces, strong metadata capture, and workflows that tolerate API or service-side changes.

In \emph{tight} co-located integration (Section~\ref{tight integration}), the HPC center can manage the end-to-end stack more directly, from scheduling and orchestration to middleware and vendor interfaces. This enables tighter alignment of libraries, container images, drivers, and monitoring/accounting with site policies for access and security. The corresponding cost is a larger operational surface area: the center must manage compatibility with rapidly evolving QC software, coordinate maintenance, and validate changes across both HPC and QC components.

\subsubsection{Latency, scheduling, and workflow structure}
Co-location can provide lower and more predictable latency between classical and quantum components, which helps workloads requiring frequent feedback (e.g., adaptive execution or iterative parameter updates) and can enable co-scheduling or reservations that couple HPC jobs and QC access. However, the QPU remains a shared resource, so there is still the need to define how QC allocations interact with the HPC scheduler.

Remote integration adds network latency and variability as well as provider-side queue dynamics, often pushing workflows toward coarser-grained patterns such as circuit batching and asynchronous submission. For many near-term applications, latency-critical classical work is modest and can be handled by a local server near the QPU (when available), while the HPC-heavy stages typically run on longer time scales where network latency is less relevant. In both models, robust production workflows generally need retries, timeouts, and back-off logic when interacting with QC services.

\subsubsection{Operational considerations}
Independent of model, HPC centers must address security, reliability, and resource governance. Tight integration concentrates responsibility locally (physical security, telemetry handling) and can enable unified monitoring, but requires local expertise and lifecycle management. Loose integration reduces local hardware operations but introduces dependencies on provider uptime, quotas, and change management, and may constrain observability to exposed telemetry. Scheduling and accounting also differ: tight integration can couple QC allocations more directly to the HPC scheduler, whereas remote integration often requires mapping HPC job lifecycles onto external submission and queue states.

\subsubsection{Maturity, scalability, and benchmarking}
Loose integration is often operationally relatively mature already today, as many QC systems are delivered as managed services with stable APIs and established support, lowering adoption barriers. Tight integration can be mature at specific sites but is more sensitive to local engineering investment and vendor integration support, and often requires adapting HPC practices to QC vendor stacks. The two models scale differently: remote access scales breadth (multiple providers/technologies without local expansion) but may be limited by external queueing and policy constraints; tight integration scales depth (lower-latency coupling and richer integration) but typically requires additional local infrastructure and staffing. 

Benchmarking should distinguish device performance from end-to-end workflow performance. Tight integration enables controlled measurement of orchestration and data-movement overheads, while remote benchmarking must account for network variability and provider queue dynamics. In both cases, comparisons are most meaningful when reporting interface mode, queue/reservation conditions, batching strategy, and full classical overheads alongside quantum metrics. It is important to compare the metrics for different integration models, including parameters such as space and utilities in addition to computational performance, that is, job throughput. A more thorough discussion on benchmarking can be found in Section~\ref{benchmarking}.


\ifSubfilesClassLoaded{%
  \nocite{apsrev41Control}%
  \bibliographystyle{apsrev4-1}%
  \bibliography{refs}%
}{}
\end{document}

\ifSubfilesClassLoaded{%
  \tableofcontents%
  \let\tableofcontents\relax%
  \setcounter{section}{3}%
}{}
\section{Potential Use Cases}
\label{use_cases}

\subsection{Challenges in Predicting Quantum Utility}
\label{use_cases_intro}

Arguably the central question for QC is, ``For which applications (if any) will quantum computers provide some sort of benefit?'' Here, benefit typically means lower cost-to-solution. Solution quality is another possible benefit, but this can almost always be recast in terms of cost-to-solution. \changes{In general, we use cost-to-solution as a shorthand for time-to-solution and/or energy-to-solution at a specified target accuracy and confidence level.} A key point is that quantum computers are not simply faster versions of conventional computers, but fundamentally different kinds of computers that can theoretically perform certain computational tasks with more favorable scaling than any classical (non-quantum) computer~\cite{Nielsen&Chuang}. Thus, while a quantum computer may provide no benefit over HPC for a small task, it may provide a substantial benefit for a large task, as schematically illustrated in Fig.~\ref{fig:fig_one} and discussed in~\cite{hoefler2023disentangling}. Finding quantum algorithms which provide some sort of definitive scaling advantage over classical (non-quantum) algorithms has been the focus of much of quantum computer science since its inception. However, quantum scaling advantage has been proven for relatively few practical applications so far, in part because the state-of-the-art in classical algorithms is constantly improving and the most effective algorithms often have better scaling in practice than can be proven. 

For this reason, inquiry has recently begun to shift toward identifying applications for which quantum computers are expected to provide \emph{utility}, i.e.\ perform useful computations at competitive cost, leaving more subtle questions of advantage to the future. But even the question of quantum utility~\cite{Herrmann2023} is non-trivial; in fact, it is currently the subject of several substantial government research programs in the US~\cite{DARPA_QB,DARPA_US2QC,DARPA_QBI,ASCR_ARQC}, and Japan~\cite{SIP3quantum}.

Thus any answers that can be given at this time must be regarded as rather tentative, subject to ongoing developments in quantum hardware, quantum algorithms, and compilation strategies.

\begin{figure}[h]
 \includegraphics[width=0.5\textwidth]{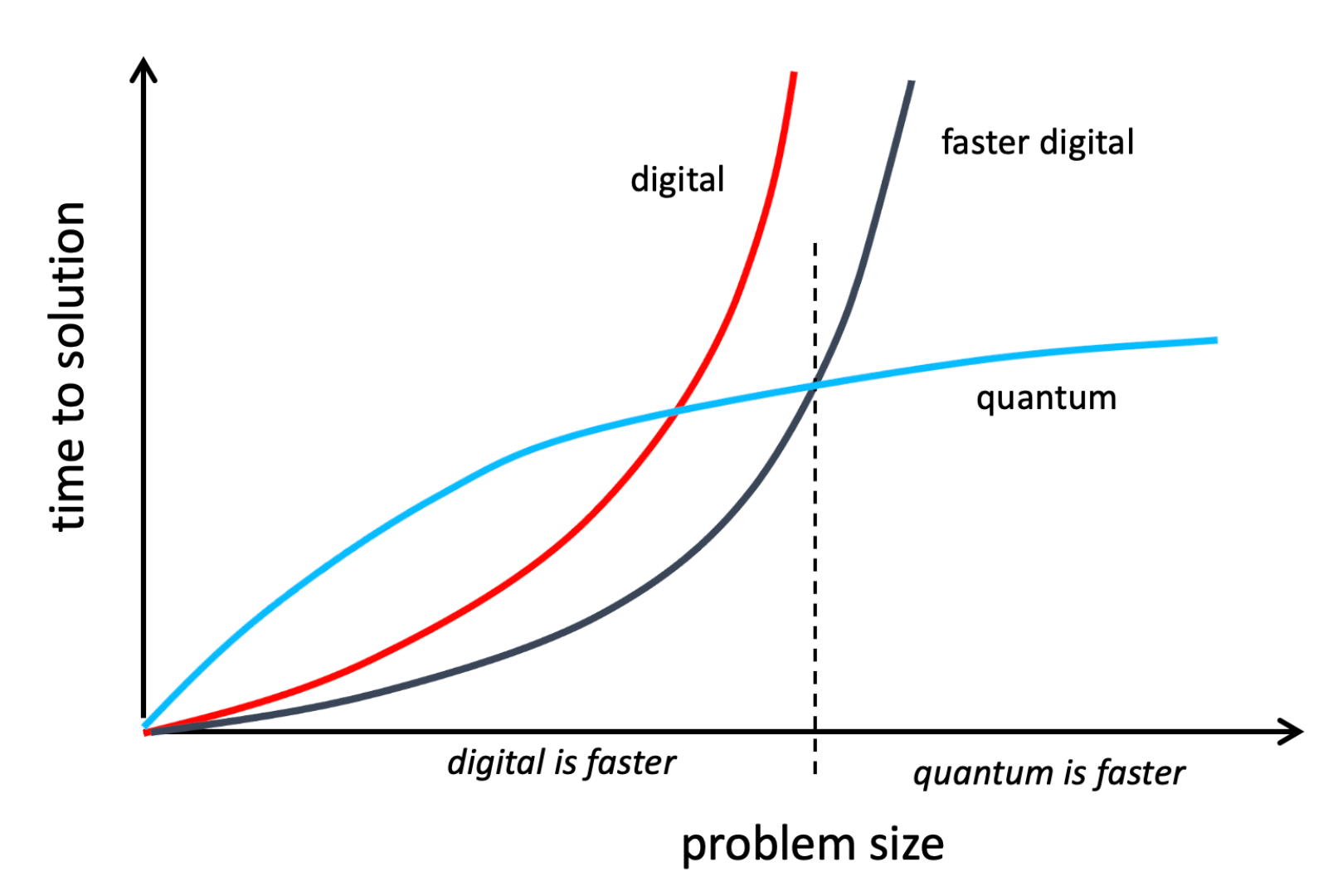}
 \caption{\label{fig:fig_one} 
 Schematic representation of the scaling characteristics of classical and quantum algorithms with problem size (axes not to scale for illustrative clarity), highlighting the potential of quantum computers and algorithms to show advantages over their classical counterparts in computing speed as problem size exceeds a certain threshold.}
\end{figure}

To understand why answering the question of quantum utility is non-trivial, it will be helpful to consider how different QC is from traditional HPC. Just as graphics processing units (GPU) introduced a new computing paradigm that required not only a reworking of existing software stacks but a rethinking of how some computational tasks were formulated, so too will quantum computers, but much more radically so.

At the algorithmic level, \changes{QC outputs for individual runs are inherently probabilistic, while the circuit evolution is deterministic.}  As introduced in \changes{Sec. \ref{QC_basics}}, quantum algorithms work by encoding and manipulating data in quantum bits or ``qubits''. The joint state of a set of qubits is in many ways analogous to a probability distribution but mathematically richer than just probabilistic bits~\cite{Chowdhury2023a} because of complex wavefunctions of quantum states.

A quantum algorithm proceeds by manipulating a few qubits at a time, progressively shaping their joint quantum state in a carefully designed way. At the end of the algorithm the qubits are measured, yielding a (classical) bit vector that is a random sample from the distribution defined by the qubits' joint state.  A successful algorithm concentrates a relatively high probability (but not necessarily all the probability) on the bit vector(s) representing the desired solution to the computational task~\cite{Nielsen&Chuang}.

Quantum algorithms often address applications by quite different approaches than conventional methods and rely on different computational primitives.  In most cases, quantum computations will not simply be drop-in replacements for computations performed in existing workflows; HPC workflows will likely have to be substantially revised to take advantage of quantum processors (QPU).  Furthermore, quantum algorithms are sometimes subject to (from the perspective of HPC) peculiar constraints on the problem parameters, or have complexity that depends on parameters that conventional algorithms are insensitive to. 

As an example of some of these novel considerations, we can mention the HHL quantum algorithm, which \changes{under restrictive assumptions can solve} the linear system $Ax = b$ exponentially more efficiently than \changes{best-known} classical algorithm.  The first caveat is that HHL is suitable only for sparse matrices $A$ with low condition number. Secondly, $b$ and $A$ are not input as data arrays, but must be encoded as efficient quantum circuits that compute $b$ and simulate $e^{iAt}$, respectively. Finally, the algorithm does not return the solution vector $x$ but allows one to obtain $x^\dagger M x$ where $M$ is the matrix representation of some efficiently implementable measurement~\cite{harrow2009quantum}. Conditions and caveats such as these often make it difficult to determine where and when quantum computational methods could provide significant benefit in practice. Indeed, later work~\cite{chia_sampling-based_2020} showed that a new classical algorithm subject to the same constraints as the HHL algorithm could recover much of the speedup it offered.

As discussed in Sec.~\ref{QC_basics}, another factor crucial to quantum utility is that, unlike classical bits, qubit states are highly sensitive to small environmental perturbations and operational inaccuracies. This sensitivity significantly limits the capabilities of current quantum devices and underscores the need for FTQC. Many \changes{QEC codes and fault-tolerance schemes} have been devised since QC was first imagined. However, substantial research and innovation are still required to reduce the resource overheads associated with fault tolerance, which must be addressed to scale quantum systems to the thousands of logical qubits necessary for \changes{many} practical applications~\cite{Beverland2022}.

In summary, a promising use case for QC first requires the identification of a valuable computational task for which there is at least one quantum algorithm that is believed to have favorable scaling. The size and complexity of a quantum computer that would be needed to perform this task then depends significantly on how the algorithm is compiled and spatially laid out, how logical operations are made fault tolerant, and the speed and reliability of the underlying quantum hardware.

In the remainder of this chapter, we will present several such potential use cases of quantum computers and provide rough estimates for the needed scale of QC based on the current state-of-the-art, keeping in mind that future improvements in hardware and algorithms could change the landscape significantly. 
Our goal is not to provide an exhaustive review but rather to provide a perspective to the HPC community. For readers seeking a deeper exploration of this topic, we refer them to comprehensive reviews and organizations roadmaps such as Refs.~\cite{Dalzell2023,alexeev2024quantum,DOE_QIS,bärtschi2025potentialapplicationsquantumcomputing} and references therein.

\subsection{Condensed Matter Physics}
\label{CMP}
Condensed matter physics (CMP) is the study of the macroscopic properties of matter using quantum mechanics to predict the collective behavior of a vast number (order of $N \approx 6.02\times 10^{23}$) of particles, such as electrons, atoms, or molecules, within investigated materials. This branch of physics addresses a large spectrum of complex problems, including phase transitions and critical phenomena, electronic properties of semiconductors and superconductors, nanomaterials, quantum magnetism, and topological materials~\cite{Kittel:2018,Modern_CMP:2019,Ashcroft_Mermin:1976}. CMP therefore underpins many technological advancements, such as developing materials with tailored properties to design next-generation electronic devices as well as quantum devices and systems such as quantum computers, sensors, and quantum networks. 

In this context, condensed matter physicists develop simplified models to study and understand the physics governing the behavior of specific materials and structures. Some models are analytically or numerically solvable in polynomial time for large scale (i.e., large number of particles) in certain limiting cases, typically in lower dimensions or under specific conditions where the complexity of interactions are limited near the extrema of spatial dimensions (very low or high dimensions) or interaction strengths (very weak or strong interactions). For general cases with arbitrary parameters, interactions, and particle densities, researchers have employed various numerical approximation methods~\cite{Fehske2008, Avella2013} to gain insights into the model's behavior and properties. However, the complexity of classically simulating problems involving strong quantum correlation effects such as \emph{superposition and entanglement} grows exponentially with the number of particles, rendering strongly-correlated systems intractable at large scale using even the most advanced HPC systems. Solving such problems is at the heart of modern CMP, with the potential to unveil and understand new emergent phenomena. It is worth mentioning that the CMP's insight on these ``twin difficulties of scale and complexity''~\cite{Anderson1972} are of relevance to the frontier of AI research as well~\cite{Wei2022}.

Following ideas proposed by Richard Feynman about four decades ago, one way to circumvent the above issue is to ``simulate quantum with quantum'' using analog quantum simulators which can be seen as special-purpose quantum computers~\cite{Feynman1982} (``not a Turing machine''~\cite{Feynman1982} so they they do not need to be capable of universal computation). It achieves this by emulating the quantum interactions that govern the system of interest, represented by a mathematical operator known as the Hamiltonian --- an observable that, when measured, reveals the system's total energy. However, analog quantum simulators are constrained by the native Hamiltonian of their physical platforms --- such as neutral atoms, ion arrays, spins in quantum dots, superconducting circuits, or photonic waveguides --- meaning they can only simulate systems governed by Hamiltonians analogous to their own~\cite{QSim_Georgescu:2014}. Although such systems are not error corrected and therefore fall in the NISQ devices category, they are already \changes{reported to} operate in a quantum advantage regime for a scientific problem, as discussed in Refs.~\cite{Daley2022,flannigan2022propagation,Trivedi2024}, and early hybrid HPC
\changes{+QC} integration initiatives based on analog quantum simulators are being implemented, e.g.~\cite{HPCQSEU}.

In principle, a quantum system's time evolution and dynamics under any condensed matter model Hamiltonian can be implemented and simulated on a digital quantum computer using universal quantum logical gates~\cite{Barenco1995} designed in analogy to a classical digital computer. There, the computation is split into discrete steps of quantum gate operations using the Trotterization technique~\cite{QSim_Georgescu:2014}. In the current NISQ era, digital quantum simulation is making significant strides, with successful demonstrations of gate-based simulations of condensed matter Hamiltonians across various platforms~\cite{Quantum_Simulators_PRX_2021}. Initially focused on small, classically tractable systems, these simulations are now approaching the boundaries of classical tractability~\cite{fauseweh2024quantum}. For example, by solving the dynamics of the transverse field Ising model (TFIM) with Trotterization, IBM's two-dimensional (2D) 127-qubit experiment~\cite{Kim2023} was the largest NISQ simulation producing accurate physical results, serving as strong evidence for quantum utility~\cite{Herrmann2023}. A flurry of classical simulations~\cite{Tindall2024, Anand2023, Liao2023, Begusic2024, Patra2024}, primarily based on various tensor network algorithms except for Ref.~\cite{Begusic2024} using Clifford perturbation theory, swiftly followed and validated the results of the quantum simulations, placing an asterisk on the claim of quantum advantage over classical methods. Notably, Patra~\textit{et al.}~\cite{Patra2024} claimed to have accurately simulated the quantum dynamics of the kicked Ising model (KIM) for even larger IBM quantum processors—the 433-qubit Osprey chip and the 1121-qubit Condor chip—on which the corresponding quantum simulations have yet to be conducted.

While the race between quantum and classical computing will continue in the battle ground of TFIM and other magnetic spin models alike before a clear winner emerges, the 2D Fermi-Hubbard model describing interacting fermions (e.g., electrons)~\cite{Editorial2013} is a time-tested hard problem for classical computers and it is proved that its ground state problem with local magnetic field is QMA-hard~\cite{Schuch2009}, i.e., a hard problem even for quantum computers. The Fermi-Hubbard model is a particularly significant model in CMP because it helps explain the microscopic mechanisms underlying various quantum materials, including high-temperature superconductors, Mott insulators, and itinerant magnets. Therefore, solving the Hubbard model at relatively large scale---say, $10{\times} 10$ 2D grid---can not only serve as a more definitive benchmark for quantum advantage (e.g., see Fig.~2 in Ref.~\cite{Daley2022}) but also provide invaluable utility in solving the most challenging puzzles in materials science. To achieve such utility on future fault-tolerant quantum computers, 200 logical qubits are required to represent all fermion modes in the $10\times 10$ 2D Hubbard model and simulate its dynamics using the Trotter algorithm~\cite{Daley2022,Dalzell2023}; for the ground state problem, additional ${\sim} 50$ logical qubits are needed to execute the quantum phase estimation (QPE) algorithm; estimated $0.1$--$100$ million T~gates are needed to simulate the dynamics and solve the ground state~\cite{Daley2022, Dalzell2023}. A considerably smaller size (8 sites encoded with 16 physical qubits) Fermi-Hubbard model was recently solved on a superconducting quantum computer~\cite{Stanisic2022}, while a system with about 80 sites was simulated with a cold-atom analog quantum simulator~\cite{Mazurenko2017}, although the latter device cannot measure arbitrary physical observables as digital quantum computers can.

\subsection{Quantum Chemistry}
\label{QChem}

Quantum chemistry is primarily concerned with the behavior of electrons in molecules and is, for that reason, often synonymous with electronic structure theory. It can serve as a powerful tool in the fundamental understanding of molecular phenomena and in industrially relevant areas such as thermochemistry and catalysis. Nonetheless, achieving quantitative results in quantum chemistry poses a substantial computational challenge due to its high demand on computational resources. Recent efforts have begun to address the application of QC to chemical problems~\cite{cao2019quantum,sajjan2022quantum,lee2023evaluating}. 

HPC has historically been an invaluable asset in computational chemistry, with the most prominent applications found in the area of electronic structure theory. In principle, once a basis set is used in the molecular orbital expansion of electronic degrees of freedom, the exact numerical answer follows from the diagonalization of the problem Hamiltonian in the space spanned by all electronic configurations afforded by the basis. This is termed Full Configuration Interaction (Exact Diagonalization in Physics circles) and is often performed in an active space (a subspace of such configurations) for computational cost reasons, and is then referred to as Complete Active Space Self Consistent Field (CASSCF). Without resorting to approximations, a space of 20 electrons in 20 (spatial) orbitals, labeled as CAS(20, 20) is already bordering the limit of what is possible in the current leadership computational facilities (LCF)~\cite{Vogiatzis2017}. The vast majority of chemistry is thus out of reach of the most adequate model.

This difficulty is hoped to be averted by turning to quantum computers. However, this is still a nascent technology and such prospects are only expected to be realized in the mature stage of fault-tolerance, when fully corrected quantum computers will be available. In the meantime, HPC is still playing a significant role since most of the leading demonstrations of QC in the chemistry space have been carried out on simulators deployed on LCFs. We highlight recent reports providing demonstrations of the variational quantum eigensolver (VQE) algorithm which is the quintessential hybrid algorithm where computational tasks are split across quantum and classical compute resources: i) with a matrix product state simulations on several hundred thousand CPUs~\cite{https://doi.org/10.48550/arxiv.2207.03711, https://doi.org/10.48550/arxiv.2303.03681}; ii) scalable general circuit simulation on leading HPC facilities~\cite{10.1145/3624062.3624221}; iii) and for industry-relevant fluoride chemistry~\cite{https://doi.org/10.48550/arxiv.2311.01242}.

These examples are still short of the postulated  $19\leq N \leq 34$ CAS($N$, $N$) size expected to exhibit a crossover between classical and quantum compute resources~\cite{https://doi.org/10.48550/arxiv.2009.12472}. In terms of QC resources, $N$ spatial orbitals, due to the Pauli exclusion principle, are fully characterized by $2N$ spin-orbitals. The simplest way to rationalize these metrics is to encode each of these spin-orbitals into a qubit~\cite{Jordan1928, Bravyi2002}, while more resource-efficient encodings can be achieved at the expense to increasing circuit complexity~\cite{Steudtner2018}. With increasing number of qubits, hardware demonstrations tend to be much less insightful in this regard because the high levels of noise make it difficult to determine if "chemical accuracy" (< 1 kcal/mol) is achievable. A recent hybrid approach enabled by HPC has pushed such chemistry applications beyond the reach of classical brute-force methods~\cite{IBM-RIKEN_2024}. However, it involves a rather simplified model of a molecule found in nature, and such heuristics lack theoretical bounds and performance guarantees, which is believed to only be achieved with FTQC. Over the years, FeMoco (iron molybdenum cofactor) has become the "holy grail" in quantum computational chemistry~\cite{Reiher2017}. Solving this problem would showcase a "real-world" application given its central role in catalyzing the nitrogen fixation reaction. To put things in perspective, the typical FeMoco CAS is (54, 54), which is more than twice as large as the largest CAS that can be classically simulated, and because the resource demand scales factorially with $N$, FeMoco is out of reach of even the most powerful HPC platforms, at least for many, many years. A fault-tolerant quantum computer is expected to simulate FeMoco in a matter of hours or days with a range of 100's to a few thousands logical qubits and $10^{14}-10^{16}$ gates acting on these logical qubits~\cite{Reiher2017, PRXQuantum.2.030305, Beverland2022, Yuri_QChem_2023, Rocca2024}, with similar analyses having been done for more modest systems~\cite{TISCC_Proceeding:LeBlond2023}. However, such estimates assume error thresholds below "chemical accuracy" or even more stringent, which is also a target unlikely to be realized in the near future. \changes{It is worth noting that recent work has advanced classical approaches by estimating the ground-state energy of a simplified CAS(76,113) FeMoco model with claimed chemical accuracy \cite{zhai2026classicalsolutionfemocofactormodel}. However, the layered complexity of FeMoco, including larger active spaces like CAS(277,404) and environmental effects, suggests that quantum advantage for a realistic chemical model in this domain might be further away than previously anticipated.}

\subsection{High Energy/Nuclear Physics}
\label{HEP}

Nuclear and high energy physics deal with a wide range of systems and complex emergent phenomena including nuclei, particle production in accelerators, nuclear reactions, supernovae explosions and the merging of neutron stars. In this context, gauge theories, which are mathematical frameworks used to describe fundamental forces and particles, are ubiquitous. Gauge theories use mathematical symmetries and transformations to explain how particles interact via force-carrying particles called gauge bosons~\cite{Peskin:1995ev}. 

The most popular gauge theories are the quantum field theories of the fundamental interactions, i.e., the standard model (SM) of particle physics. While certain immediate forecasts of these field theories can be determined using perturbative methods, most processes of interest necessitate non-perturbative simulations. These involve configuration spaces, or Hilbert spaces, with dimensions exceeding the total number of atoms in the universe.

In particular, the calculation of the interactions between quarks and gluons is computationally demanding for classical computers due to several factors. Quantum chromodynamics (QCD) is a non-Abelian gauge theory -- the force between quarks intensifies as they move apart, a phenomenon known as confinement, leading to complex interactions that are hard to compute~\cite{Gattringer:2010zz}. The particles involved, quarks and gluons, exhibit a high degree of entanglement, requiring vast computational resources to model accurately. Furthermore, the self-interaction of gluons in QCD adds a large number of degrees of freedom to the system, thus increasing its complexity. The theory operates across a wide range of energy scales, from the low energies of hadronic physics to the high energies of quark-gluon plasma. 

Lattice QCD, a standard numerical technique where space-time is discretized, faces the "sign problem", where integrals become highly oscillatory and challenging for numerical methods~\cite{Kronfeld2022, Davoudi2021}. Different ideas have been pursued to overcome such sign problems~\cite{Alexandru2022,Nagata:2021ugx}, but these problems are believed to be NP-hard~\cite{Troyer2015}. Lattice QCD relies on HPC and advanced software to provide precision calculations of the properties of particles that contain quarks and gluons. In HPC systems, the uniform space-time grid can be divided among the processors of a parallel computer. Some recent estimates on cost for the most expensive B-physics calculations at Cori (NSERC) and Summit (ORNL) require 1.5 Exaflop hours for lattice volumes of $32^{3} \times 64$ and 1500  Exaflop hours for volumes of $128^{3}\times 512$~\cite{boyle2022lattice}.

Quantum computers hold significant promise for lattice QCD calculations due to their ability to efficiently simulate quantum mechanical systems, potentially overcoming challenges such as the sign problem and handling exponentially large Hilbert spaces. However, progress is needed to make them suitable for practical lattice QCD applications. For example, developing efficient methods to encode the SU(3) gauge Hamiltonian—a fundamental component of QCD—could help address the current complexity associated with high qubit overhead and the need for advanced quantum gates. Similarly, innovative approaches to reconcile the continuous nature of gauge fields in lattice QCD with the native discrete states of quantum computers are essential to minimize truncation-induced systematic errors. These advancements would mark significant steps toward realizing the potential of QC for QCD.

Furthermore, in a similar manner to the CMP and quantum chemistry problems described earlier, the resource estimation for state-of-the-art methods reveals a demand for the number of physical qubits, gate operations, and error-correction overhead that significantly exceeds the capabilities of current quantum hardware. For example, the gate count scales as $d\Lambda t^{3/2}(L/a)^{3d/2}\epsilon^{-1/2}$ for the SU(2) and SU(3) lattice gauge theories in the irreducible representation basis, when approximating the time-evolution operator via Trotterization for a maximum error $\epsilon$ for an arbitrary state. Here, $d$ denotes the dimensionality of space, $\Lambda$ is the gauge-field truncation in the irrep basis, $L$ is the spatial extent of a cubic lattice and $a$ denotes the lattice spacing. For an accuracy goal of $\epsilon = 10^{-8}$ and a lattice with tens of sites along each spatial direction, the simulation requires $\mathcal{O}(10^{11})$ qubits and $\mathcal{O}(10^{50})$ T-gates~\cite{Shaw2020,kan2022lattice}. Recent studies estimating the resource requirements for the simulation of nuclear effective field theories (EFTs) report a T-gate count of $4\times10^{12}$ and 10,000 logical qubits for simulating a compact Pionless EFT~\cite{watson2023quantum}. 

It is also worth noting that the Schwinger model, which describes quantum electrodynamics (QED) in 1+1 dimensions and retains key features of QCD such as confinement and a topological theta vacuum~\cite{Schwinger_1962}, serves as a valuable testbed for quantum simulations. While the continuous Schwinger model is exactly solvable, its lattice version introduces complexities that make classical approaches intractable for large system sizes. The model has been experimentally demonstrated on current NISQ quantum processors for both ground state preparation and real-time evolution (see e.g. Refs.~\cite{Schwinger_2022,Schwinger_Innsbruck_2019}). 
Recently, the vacuum of the lattice Schwinger model was prepared on up to 100 qubits on one of IBM's largest quantum processors~\cite{Farrell2023}, and hadron dynamics was simulated using 112 qubits~\cite{farrell2024quantum}.

\subsection{Cryptography -- Factorization}
\label{Crypto}

Cryptography, the practice and science of securing information from being intercepted by third parties, is at the heart of all modern communications. Typically, cryptographic systems (referred to as ``cryptosystems'') are used to encrypt information that a \textit{sender} wishes to transfer, often along a public communication channel, to a \textit{receiver}. The security of the data relies on the fact that even though it may be successfully intercepted by an eavesdropper, only the receiver is able to decode it (at least in an ideal world), thus extracting its meaningful information content.

Cryptographic protocols are often divided into public-key, secret-key, and hash-function-based (for information on the latter two see, e.g.,~\cite{jirwan2013review,al2020review}. Many of the cryptosystems of today utilize a public-key based approach, where two different \textit{keys} (one public, and the other private) are used for encrypting and decrypting information. Public-key cryptosystems (PKCs) typically rely on mathematical one-way-functions; ones that are easy to compute but whose inverse is in turn difficult to calculate. The security of the RSA~\cite{rivest1978method} algorithm, one of the most utilized by PKCs, for example, relies on the perceived inability of the eavesdropper to factor large numbers into a product of primes. This is a result of the fact that best-known classical factoring algorithms (number field sieve), have runtimes that scale sub-exponentially in the number-size, \textit{i.e.} $2^{\mathcal{O}(b^{1/3})}$, where bit length $b = [\log_2(n)] + 1$ for an integer $n$~\cite{ha2022resource}. Through the use of modern HPC systems, researchers have been able to tackle problems with key sizes of 795 bits (RSA-240, i.e., 240 decimal digits) or 829 bits (RSA-250), at a runtime cost of approximately 12-days of cluster compute time for RSA-240 (equivalent to ${\sim}900$ core-years) and ${\sim}2700$ core-years for RSA-250, respectively~\cite{boudot2020comparing,Dalzell2023}. 

In comparison, factoring a medium size RSA-like number with ${\sim} 200$ bits typically takes a few core-seconds\footnote{This can be tested on a SageMath web server (\url{https://sagecell.sagemath.org}) with the code: \texttt{time factor(random\_prime(2\^{}100) * random\_prime(2\^{}100))}.}. It is believed, however, that unless substantial algorithmic advances are made, even the most powerful classical supercomputers of tomorrow will struggle when larger key sizes are used (e.g., 2048 bits or beyond in the case of RSA). 
\\
Additionally, it is worth pointing out that other PKCs have also been proposed and are sometimes used. These include Elliptic Curve Cryptography (ECC)~\cite{koblitz1987elliptic} based on elliptic curve logarithm problems, or the Digital Signature Algorithm (DSA) which is based on modular arithmetic in a finite field (see e.g. Ref.~\cite{schneier2007applied}). Although both of these example approaches require shorter keys to achieve equivalent-to-RSA bit-security\footnote{The bit-security, a metric that describes the security level of a given cryptosystem, does not in general coincide with the size of the key that it may use, see e.g. Ref.~\cite{scholten2024assessing} for details.} (thus having lower communication bandwidth requirements) they have not gained as broad an acceptance as the RSA-based approach. Most importantly however, just like the RSA-based PKCs, their security can be severely compromised if one envisions using non-classical computation, as outlined below.

In 1994, Peter Shor discovered a factoring algorithm that can run efficiently on a quantum computer~\cite{shor1994algorithms}, thus fundamentally changing how the future security of widely used PKCs should be treated.
It is crucial to point out that the Shor's factoring algorithms not only affects the security of RSA-based systems, but also other PKCs (such as ECC or DSA), through its ``discrete logarithm'' variant ~\cite{Dalzell2023,scholten2024assessing}. 

While only proof-of-concept implementations of Shor's algorithm have been explored in the current NISQ era—such as the demonstration of factoring the number "21" in~\cite{skosana2021demonstration}—it is expected that large, fault-tolerant quantum computers will be necessary to fully break today's RSA keys (2048-bit or even 1024-bit). Nonetheless, the advancements made over the past three decades have significantly influenced how governments and security agencies perceive this technology.

Various estimations of the QC resources needed to make the Shor's algorithm practically useful have been explored. For instance, a surface-code-based architecture may require approximately 1 to 20 million physical qubits, with corresponding runtimes ranging from about a week to several hours ~\cite{Gidney2021howtofactorbit,gidney2025factor2048bitrsa}. These types of studies make many assumptions related to underlying hardware, error models, and even algorithmic details, and thus may results is somewhat different predictions. However, it is evident that as quantum devices become more sophisticated—which may take many years—breaking encryption that is currently considered secure could become feasible.

It is crucial to point out, however, that even when realized at large-scale, fault-tolerant quantum computers are not expected to be a complete solution for breaking data security. Indeed, in recent years, several post-quantum classical algorithms have been proposed~\cite{chen2016report} that are designed to be secure against both quantum and large-scale HPC attacks. Additionally, fully quantum-based cryptosystems relying on quantum key distribution (QKD) have also been studied and even realized - see e.g. Refs.~\cite{bourgoin2015experimental,chen2020sending}. QKD relies on transmitting quantum states between parties and, under ideal conditions, guarantees the detection of eavesdroppers, enabling the parties to ascertain whether data has been compromised, all in accordance with the principles of quantum mechanics. The security of these systems depends on loophole-free implementation of the underlying hardware, and not on clever algorithms or available computing resources. See~\cite{scarani2009security,xu2020secure,renner2023quantum,Renner_QKD_2022} for details. 

In summary, the complete role of HPC in compromising the security of cryptosystems may not yet be fully understood. On one hand, as classical compute resources continue to \changes{grow}, they will be able to decode older data sets, encoded with traditional PKCs and smaller key lengths. But as key lengths increase (as dictated by government agencies such as NIST or the NSA - see Ref.~\cite{scholten2024assessing} for a historical overview of those changing recommendations), or as either classical or quantum post-quantum cryptography-based approaches take over, the direct impact \changes{of HPC} may become more limited. This prognosis may be affected, however, by breakthrough algorithms that have not yet been discovered.

\subsection{Combinatorial Optimization}
\label{Optim}

Optimization is a common goal or subroutine in many scientific programming workflows. Perhaps the simplest conceivable optimization problem is unstructured search for a single solution from amongst a pool of $M$ candidates.  A conventional algorithm requires time scaling linearly in $M$ to solve this problem---because the search is unstructured, a conventional computer has no better way to solve it than by checking candidates until the correct solution is found. Remarkably, a quantum algorithm is provably more efficient. Indeed, Grover's search algorithm requires a number of steps scaling as $\sqrt{M}$, a quadratic improvement relative to conventional methods~\cite{grover1996fast}. This and similar findings have inspired research into potential quantum advantages in broader instances of optimization. 

Beyond the unstructured search example above, optimization algorithms have diverse applications in applied mathematics, science, and engineering.  Examples include variationally solving ground-state problems for physics or chemistry as well as efficiently allocating limited resources in logistics. Here we focus on combinatorial optimization, which is targeted in certain HPC workflows~\cite{eckstein2015pebbl,eckstein2001pico,crainic2006parallel,langer2013parallel,kirkpatrick1983optimization,fang2014parallel,lulli2015highly} as well as quantum algorithms described below. HPC and quantum algorithms also target continuous optimization, including recent formulations of quantum-enhanced gradient descent~\cite{leng2023quantum,rebentrost2019quantum,kerenidis2020quantum}; for further discussion of related ideas we refer the reader to Ref.~\cite{abbas2024challenges}. 

Combinatorial optimization problems seek to identify an $N$-bit string $\bm z = (z_1, z_2, \ldots, z_N)$ that minimizes a cost function $C(\bm z)$. A canonical example is the travelling salesperson problem, where the target path minimizes the distance or expense in travelling between a set of cities, and $z_i$ represent decision variables for different potential paths.  Combinatorial optimization problems often belong to the NP-hard computational complexity class, meaning they require computational resources that scale exponentially in the problem size in the worst case. This motivates intensive research to improve efficiency and scalability of classical ~\cite{eckstein2015pebbl,eckstein2001pico,crainic2006parallel,langer2013parallel} or quantum~\cite{abbas2024challenges} algorithms. Approximation algorithms or heuristics are often employed when exact solutions are computationally intractable~\cite{kirkpatrick1983optimization,fang2014parallel,lulli2015highly,goemans1995improved}. 

A couple of decades ago it was realized that adiabatic evolution of a quantum state could, in principle, produce optimal or approximate solutions to optimization problems~\cite{farhi2000quantum,farhi2001quantum}.  This motivated the development of today's commercially-available quantum annealer devices that can address problems with thousands of variables $z_i$, where each variable maps to a qubit in the annealer~\cite{yarkoni2022quantum}.  Ideally, these utilize quantum effects in the optimization, though in practice the technologies are limited by noise, and the extent to which quantum effects actually play a role is a topic of ongoing debate~\cite{albash2015reexamining,king2022coherent}.  Nonetheless, these devices, as well as related Rydberg-atom analog devices~\cite{ebadi2022quantum,wurtz2022industry}, are currently capable of approximately solving large problem instances that could today be incorporated into a scientific computing workflow~\cite{JulichPasqal2023}.

More recently, digital quantum computers have become available, creating opportunities to experimentally implement algorithms such as the quantum approximate optimization algorithm (QAOA), which has been proposed as a promising approach for solving combinatorial optimization problems on near-term quantum devices~\cite{farhi2014quantum}. A number of encouraging results have been obtained for QAOA, including theoretical solutions that outperform semi-definite programming for the Sherrington-Kirkpatrick spin glass problem~\cite{farhi2022quantum}, theoretical solutions that outperform conventional algorithms for ``large-girth" MaxCut instances~\cite{basso2021quantum}, and an asymptotic scaling advantage observed in numerical simulations of QAOA solving the low-autocorrelation binary-sequence problem~\cite{shaydulin2023evidence}.  There have also been a variety of interesting studies of relations between problem structure, depth, and performance~\cite{lotshaw2021empirical,herrman2021impact,lots2023simulations,lotshaw2023approximate,lotshaw2022scaling,akshay2020reachability,wurtz2022counterdiabaticity,diez2023quantum,crooks2018performance}, advanced circuit embeddings~\cite{moondra2024promise,herrman2021globally}, heuristics for improved circuit efficiency~\cite{shaydulin2023parameter,he2024parameter}, implementations for constrained optimization~\cite{goldstein2024convergence,hadfield2019quantum}, noisy performance analyses~\cite{lotshaw2023modeling,quiroz2021quantifying,lotshaw2024exactly,xue2021effects,wang2021noise}, proposed algorithmic improvements~\cite{herrman2022multi,zhu2022adaptive,tate2023warm}, and attempts to improve or utilize QAOA by combining it with conventional algorithms~\cite{ponce2023graph,maciejewski2024multilevel,ushijima2021multilevel,dupont2024extending,brady2023iterative,wurtz2024solving}. Promising alternatives to QAOA have also been presented, including quantum imaginary time evolution~\cite{morris2024performant}, quantum interior point methods~\cite{augustino2023quantum}, and algorithms based on gene expression programming~\cite{alvarez2023gene}; related ML based approaches also exist for chemistry applications~\cite{nakaji2024generative}.

While QAOA displays known advantages for a couple of problem types, these still remain somewhat removed from typical practical applications, and like many real-world quantum applications, their resource requirements remain far from the reach of today's devices~\cite{basso2021quantum,shaydulin2023evidence,lykov2023sampling}. Also, the scaling of resources for circuits that achieve high performance in more practically useful problems, at large sizes, remains largely uncertain. This uncertainty is largely due to the communities inability to evaluate QAOA on large cases, as well as an absence of mathematical proofs on performance for generic cases. Further testing and improved quantum technologies appear necessary to understand realistic resource expectations for QAOA. 

For the HPC community, there are opportunities to design and implement workflows that can benefit from quantum optimization algorithms. A real-world optimization use-case leveraging QAOA is the optimization of metamaterial design using an active learning algorithm~\cite{kim2024performanceanalysisoptimizationalgorithm, alexeev2024quantum}. Here, combining AI, QAOA, and HPC enables efficient exploration of the complex parameter spaces in metamaterial optimization. While AI/ML creates surrogate models to predict how a metamaterial's geometry affects its performance, a variation of the QAOA, the Distributed Quantum Approximate Optimization Algorithm (DQAOA)~\cite{kim2025distributedquantumapproximateoptimization}, optimizes these models to identify designs meeting desired criteria. In this context, HPC accelerates simulations of optical properties and performance metrics. This iterative approach streamlines the design of high performance metamaterials and highlights the power of merging AI, QC, and HPC for large-scale optimization in materials science (see section \ref{AI} for more examples of the combination of AI and QC).

There may be additional applications where combinatorial optimization can serve as a useful subroutine of a larger application, and can benefit from quick heuristic solutions such as those provided by quantum algorithms. We expect that successfully combining quantum and HPC resources will be especially important in next generation HPC workflows.

\subsection{Artificial Intelligence}
\label{AI}

Artificial Intelligence (AI) and Quantum Information Science (QIS) represent the forefront of cutting-edge modern science, each driving revolutionary advancements in science and technology. QC is poised to elevate certain aspects of ML, such as linear algebra routines, but the synergy between QIS and AI is much deeper and advancements in one domain could reciprocally fuel progress in the other~\cite{Castelvecchi2024}.  Fully harnessing this mutually reinforcing interaction is a crucial step toward realizing the full potential of quantum technologies (e.g. in scalability, cost-effectiveness, and applicability). In this section, we discuss how these two different paradigms can cooperatively evolve, rather than considering these two disciplines to be in competition~\cite{MIT_AI_QC}. 
\changes{Of particular relevance}, we discuss the field of QML that has evolved rapidly over the past two decades ~\cite{QMLreview2017,QML_2024comprehensive}.

\subsubsection{Quantum Machine Learning (QML)}
\label{QML}

The field of QML covers many use cases~\cite{sajjan2022quantum,huang2023learning}, often delineated by the nature of the input data, whether classical or quantum, and the specific algorithm employed, spanning supervised, unsupervised, or reinforcement learning paradigms.

Near-term QML in the NISQ era benefits significantly from the integration of classical and quantum resources across several key areas~\cite{babu2025entanglement}. For instance, variational algorithms leverage HPC capabilities to enable parallelization of data and tasks, as well as the utilization of accelerators. Quantum-specific tasks include advanced techniques such as mid-circuit measurement~\cite{mid_circuit_meas_2019} and circuit knitting~\cite{circ_knitting_2025}, while classical tasks involve strategies like distributing multiple quantum learners in ensemble models~\cite{Q_Ensembles_2024} or performing batched evaluations of gradient terms~\cite{Q_gradients_2024}.

Long-term QML applications are poised to benefit from the acceleration and distribution of quantum tasks. For instance, dynamic circuits—where the quantum circuit adapts in real-time based on classical inputs—highlight the role of HPC as an accelerator. These circuits have been shown to achieve algorithmic advancements, such as in quantum phase estimation (QPE), where classical processing of auxiliary qubit measurements with feedback improves circuit performance~\cite{Corcoles2021}. Additionally, QML could leverage FTQC algorithms as subroutines, such as the HHL algorithm or distance estimation techniques, to enhance functionality and scalability.

Designing and deploying quantum analogues of classical ML models and methods have produced many near-term QC for AI applications.  However, na\"{i}vely incorporating variational quantum circuits into data-heavy supervised training workflows immediately encounters bottlenecks associated with the serial processing of training samples while further amplifying these issues through the high precision requirements of data encoding and output retrieval. Establishing a quantum advantage with variational NISQ algorithms is difficult, as it heavily depends on the training data and involves numerous subtleties and ambiguities in identifying suitably challenging problems. For example, the V-score benchmark~\cite{wu2024variational} can be used to identify many-body problems that will be challenging to approximate using classical variational methods.  In contrast, establishing such a score to identify challenging problems in classical ML remains an open question. Thus, there is a growing sentiment within the QML community to shift focus from pursuing the elusive goal of quantum advantage in learning tasks toward demonstrating practical quantum utility—a perspective increasingly reflected across the broader field of QC, as discussed in Sec.~\ref{use_cases_intro}.

Identifying problems well-suited to quantum computers can lead to potential paths towards quantum utility. Generative modeling, a versatile approach that encompasses tasks like fitting data distributions and serving as surrogate models, represents a promising area. While recent results in quantum statistical learning theory suggest that the complexity associated with training quantum generative models does not guarantee quantum advantage~\cite{arunachalam2020quantum}, these models offer a practical edge by reducing the computational burden associated with computing normalization factors or partition functions.

Another potential path to utility could be found in the areas of training on small datasets, and transfer learning. Rather than attempting to replace large classical AI models with fully quantum instantiations, the focus could shift to using QC and QML models to enhance the efficiency of existing models through parameter pruning~\cite{Yuri_LLM_2024}.

\ifSubfilesClassLoaded{%
  \nocite{apsrev41Control}%
  \bibliographystyle{apsrev4-1}%
  \bibliography{refs}%
}{}
\end{document}

\ifSubfilesClassLoaded{%
  \tableofcontents%
  \let\tableofcontents\relax%
  \setcounter{section}{4}%
}{}
\section{Quantum Emulation}
\label{emulation}

\subsection{Present and future benefits of classical emulation}
\label{emulation_benefits}
Classical emulators play a crucial role in the development, testing, and analysis of quantum circuits, algorithms, and states. They also provide an environment for exploring quantum noise and errors in a controlled manner. Despite the immense potential of quantum computers, classical emulators remain an indispensable component of the QC toolkit, driving progress and innovation in the field. Additionally, these emulators are likely to continue serving as circuit execution environments within multi-node compute clusters, handling distributed workloads effectively~\cite{alexeev2023quantumcentric}. \changes{In this section, we use ``emulation'' as an umbrella term for classical execution of quantum programs, including both ideal circuit simulation and noise-aware simulation.}

Quantum computers are typically rare and in high demand, whereas classical (super)computers are generally more accessible. The development of quantum circuits and algorithms requires numerous trials and errors, an iterative process that can be significantly accelerated through emulation, avoiding long wait times for scarce quantum resource and providing a debugging interface similar to classical programming workflow.

Currently, quantum computers are prone to noise, and FTQCs may still be decades away. In contrast, emulated circuits can be either noise-free or configured with adjustable noise levels, which is crucial for developing algorithms suited for future FTQCs as well as for those designed to operate on today’s noisy quantum computers.

Some QPU platforms, such as ion traps and neutral atoms in optical tweezers, allow for the flexible individual positioning of ions or atoms carrying qubits, enabling reconfigurable entanglement and arbitrary connectivity. In contrast, most other QPU platforms, such as superconducting qubits, typically have a fixed qubit connectivity topology defined by the hardware design. Although quantum circuits can be transpiled to fit specific qubit topologies, this often results in a significant increase in gate count due to the need for SWAP gates, which can also reduce overall fidelity. In contrast, emulated quantum computers can accommodate a wide range of topologies, making it possible to emulate quantum circuits and algorithms across various QPU configurations on the same classical hardware.

\changes{\subsection{Classical emulation for the certification of quantum advantage}}
\label{certification}
There have been multiple claims of quantum advantage~\mbox{\cite{arute2019quantum,Wu2021,Zhong2020,Zhong2021,Madsen2022}} (also known as quantum supremacy/primacy~\cite{Durham2021}), in which a QC can perform a task at a significantly lower cost than classical computers. This \changes{comparison} depends not only on the quantum computer used, but also on the classical \changes{emulation methods or }algorithm 
\changes{benchmarked against it}. For example, as discussed in Sec. \ref{CMP}, Ref.~\cite{Kim2023} argued that a noisy 127-qubit machine successfully estimated accurate expectation values which are beyond brute-force classical calculations. However, the authors did not rule out the possibility of sophisticated classical algorithms outperforming their QCs for the problem, and in fact the quantum \changes{emulation} results were very soon reproduced and even surpassed (in terms of accuracy) by multiple subsequent studies using classical \changes{emulation} methods (e.g.,~\cite{Tindall2024,Anand2023,Liao2023,Begusic2024,Patra2024}). In the same context, the original ``quantum supremacy'' experiment~\cite{arute2019quantum} had also been solved classically a few years later~\cite{Pan2022a}.
To accurately evaluate the capability of QCs, classical \changes{emulation} methods need to be constantly updated as QCs evolve~\cite{Xu2023}. \changes{Emulation tools are invaluable for reproducing, validating, and surpassing results, allowing rigorous benchmarking of quantum and classical systems.} Quantum computers are expected to take a long time to surpass classical computers on many conventional problems that have been efficiently solved classically. Therefore, the primary target of quantum advantage should be quantum-oriented not only at present but also in the future. Because quantum emulators would be suitable for such problems, they will be important in certifying any claimed quantum advantage.
\\
\subsection{Methods to emulate QCs}
\label{emulation methods}
Several methods have been developed to emulate QPUs, each offering distinct advantages and disadvantages. The most popular approaches are introduced below and summarized in the accompanying Table \ref{tab:tab_emulation_summary}. \changes{Note that in the Table, ``Noise'' refers to whether the emulator can incorporate an explicit noise model (e.g., depolarizing, amplitude damping, device-calibrated channels); it does not imply that tensor-network methods are inherently ``noisy'' or that full-state methods must be ``noiseless''.} For readers seeking a deeper exploration of QPU emulation methods, we refer them to recent reviews such as Refs.~\cite{Alexeev2023,Xu2023} and references therein.

\begin{table*}
\caption{Popular and general-purpose QPU emulation methods.} 
\label{tab:tab_emulation_summary} 
\begin{tabular}{| p{0.14\linewidth} | p{0.15\linewidth} | p{0.15\linewidth} | p{0.14\linewidth} | p{0.1\linewidth} | p{0.2\linewidth} |}
\hline
\multicolumn{2}{|c|}{Method} & Emulation example & Accuracy & Noise & Target \\
\hline
\multirow{4}{*}{Full state}
 & \multirow{2}{*}{State Vector}   & 48 qubits~\cite{raedt2019massively} & \multirow{2}{*}{Exact} & Noiseless  & Small circuits \\
 &                                 & Sunway TaihuLight and K &                        & or approx. & (both shallow and deep)  \\
\cline{2-6}
 & \multirow{2}{*}{Density Matrix} & 21 qubits~\cite{li2024tanq} & \multirow{2}{*}{Exact} & \multirow{2}{*}{Both} & Small circuits \\
 &                                 &  NERSC Perlmutter  &                        &                       & (both shallow and deep)  \\
\hline
\multirow{4}{*}{Tensor network}
 & \multirow{2}{*}{MPS}  & 1000 qubits $\times$ 100 depth~\cite{sun2023improved} & \multirow{2}{*}{Approximation} & \multirow{2}{*}{Noisy} & Shallow circuits \\
 &                       & >250 Fugaku nodes &                       &                       & (limited to simple entanglements) \\
\cline{2-6}
 & \multirow{2}{*}{PEPS} & 100 qubits $\times$ 42 depth~\cite{liu2021closing}    & \multirow{2}{*}{Both}          & \multirow{2}{*}{Both}  &  Shallow circuits \\
 &                       & 107k Sunway nodes &                       &                       & (including 2D systems) \\
\hline
\end{tabular}
\end{table*}

\vspace{1em}
\subsubsection{Full state methods}
\label{ssec:full_state_methods)}

The most straightforward way to emulate quantum circuits is to hold all state vector components of emulated qubits in memory and apply operations to them. This method, commonly referred to as the state vector method, is suitable for deep circuits with many ideal, i.e. noiseless gates and convoluted entanglements. 

While circuit noises can be approximated to some extent in the state vector method, an exact representation of different types of noises requires an extended format of the quantum state. This can be achieved with mixed state emulations that express states as density matrices~\cite{qsim} and are capable of exact modeling of noise sources, at the cost of increased resource usage.

Emulating a generic quantum circuit in full-state is exponentially costly in time and/or memory, limiting its application to small circuits. A $n$-qubit quantum register requires $2^{n}$ array vectors containing amplitudes for the full state. Since each amplitude is a complex number requiring two 8-byte double data types representing a real and an imaginary part, the total memory size for the n-qubit quantum register reaches $2^{n+4}$ bytes. Emulating a density matrix requires significantly more memory than emulating a state vector because the density matrix is a $2^{n} \times 2^{n}$ complex-valued matrix. For instance, describing the idealized state vector of a 53-qubit Sycamore~\cite{arute2019quantum} processor would require around 144 petabytes, and a density matrix description for its true, noisy behavior would demand $\sim 1.3 \times 10^{18}$ petabytes~\cite{jones2023distributed}. To put this in perspective, this is roughly $\sim 10^{10}$ times more than the amount of information generated by humankind in 2023, which was estimated to be $120$ zettabytes~\cite{Worlddata_2023}. Serial emulation of even simple quantum operations on such large states is therefore impractical due to its extreme slowness. \changes{This comparison is only meant to convey scale: full-state methods quickly become memory-bound at leadership scales, which is why distributed state-vector emulation and tensor-network methods are essential in practice.}

For state vector emulators, classical emulation methods are highly efficient at parallelization, effectively utilizing classical hardware acceleration techniques such as multithreading and GPU-based parallel processing. Nonetheless, finite memory limitations require distributing the state vector across multiple machines, which collaborate over a network to store manageable subsets of the full state description. This approach enables the study of larger quantum systems by increasing parallelization, facilitating more efficient emulations. Under ideal weak scaling, 1024 compute nodes of a supercomputer could collectively emulate a 40-qubit quantum state in a time comparable to a single node emulating a 30-qubit state~\cite{jones2023distributed}. This distribution approach is therefore essential for handling large data structures and represent a common use of supercomputing platforms.

In summary, developing emulators for distributed and parallel systems requires advanced skills in programming and algorithm design. Challenges arise due to the absence of globally shared memory and synchronized program clocks, which are standard in serial computing. Performance metrics and runtime predictions must account for inter-node communication costs, as data exchange over a network incurs significant latency compared to local operations, even on HPC networks~\cite{motlagh1998memory}. Thus, optimizing communication becomes crucial, especially for memory bandwidth-bound state vector emulations~\cite{haner2016high}. Designing distributed algorithms for emulating quantum operators, even in an unoptimized state, thus presents substantial complexity. 
A more comprehensive discussion of full-state methods can be found in Ref.~\cite{alexeev2024quantum} and references therein.

\subsubsection{Tensor network methods}
\label{tensor networks}
Quantum gates can be represented as tensors acting on state vectors. Hence, quantum circuits can be regarded as a network of tensor products applied to initial qubit states. Multiple methods have been developed to contract the tensors efficiently without expanding the full state, thus relaxing memory requirement (Fig. \ref{fig:tensor_network_methods}). Such "tensor network methods" can handle a larger number of qubits, provided the circuit remains sufficiently shallow and an approach suited to the entanglement structure is applied.

\begin{figure}[h]
 \centering
 \includegraphics[width=0.45\textwidth]{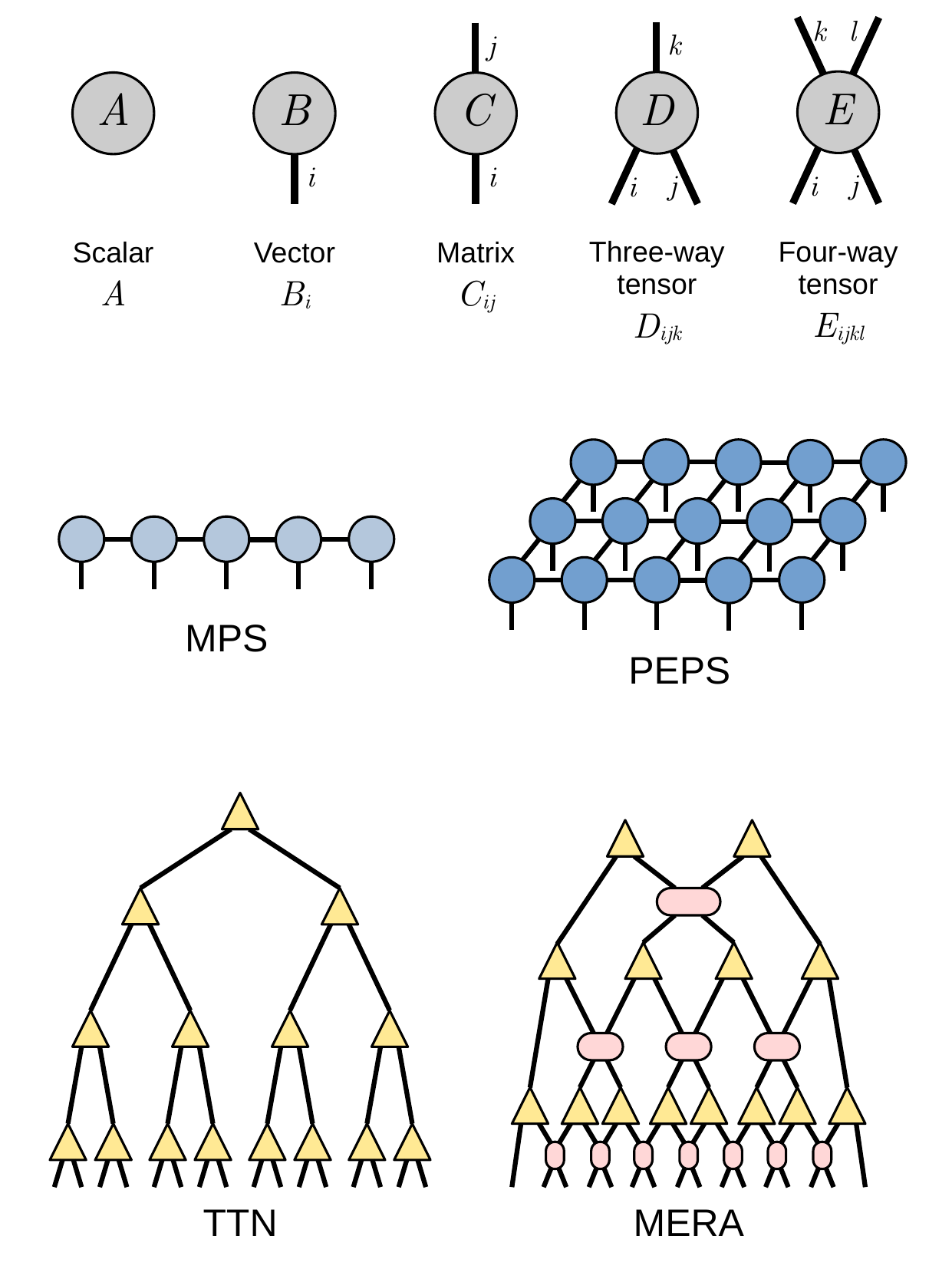}
 \caption{Architectures of tensor network methods. Tensors are represented as circles or triangles with multiple legs. MPS is optimal for one-dimensional systems with local interaction. PEPS is suitable for multiple dimensions. The tree-like structure of TTN makes it suitable for systems with hierarchical entanglements. MERA contains ``disentanglers'' (pink circles) reducing short-range entanglement, and is suitable for systems with multiple length scales. \label{fig:tensor_network_methods}} 
\end{figure}

In the matrix product state (MPS) method, quantum circuits are represented as tensors in a one-dimensional chain. \changes{This can reduce the effective cost dramatically for circuits with limited entanglement; the runtime and memory typically scale polynomially in the number of qubits and in the bond dimension (which grows with entanglement), but can revert to exponential behavior in the worst case.} Note that this memory efficient technique can only \emph{approximate} the quantum state, with the error increasing as the entanglement structure becomes more complex (e.g. adjacent entanglement among two-dimensional qubits).

Multiple representations of tensor networks have been devised to better handle complex circuits. The projected entangled pair states (PEPS) is an extension of MPS to higher dimensions, making it suitable for high-dimensional circuits but computationally more expensive than MPS. Other types of tensor network emulators include tree tensor networks (TTN), multi-scale entanglement renormalization ansatz (MERA), and continuous variable tensor network (CVTN). For readers seeking a deeper
exploration of tensor network methods, we refer them to comprehensive reviews such as Refs.~\cite{banuls2023tensor,evenbly2022practical,orus2014practical} and references therein.

\subsubsection{Other emulation methods}
\label{other emulation}
While full state methods and tensor network methods are widely used for general purposes, other types of emulation methods have been devised as well. Some methods are tailored for stabilizer circuits, which are important for quantum error correction~\cite{seddon2021quantifying}. 
\changes{Clifford algebra-based emulators, in particular, are highly efficient for simulating stabilizer circuits, as described by the Gottesman-Knill theorem~\cite{aaronson2004improved}. This theorem demonstrates that quantum circuits composed entirely of Clifford gates (along with Pauli measurements and certain initial states) can be simulated efficiently on classical hardware. These emulators are powerful tools for studying quantum error correction codes and related tasks but are not universal, as they cannot, for example, simulate quantum computations involving non-Clifford gates.} The decision diagram~\cite{samoladas2008improved} and tensor diagram~\cite{hong2022tensor} methods exploit redundancies in the quantum circuit in graph-based approaches. Variational algorithms for QCs contain repeated circuit structures with different gate parameters. As an example, Huang et al.~\cite{huang2021logical} compiled such circuits to logical formulae that support efficient sampling for different gate parameters, boosting circuit sampling oriented for variational algorithms.

In Summary, the benefits of emulation stem from the flexibility, abundance, and noise-controllable nature of classical computers. Different emulation methods have different limits to targetable qubit sizes, circuit depths, QPU topologies, etc. The state vector method can handle deep circuits and all-to-all connections, but is limited to a few dozens of qubits. On the other hand, tensor network methods are unsuitable for deep circuits involving convoluted entanglements, while they can handle large number of qubits as long as the circuit is shallow. Regardless of the method used, emulation of large or deep quantum circuits presents a significant challenge for HPC systems. This is natural, as the advantage of QC comes from the ability to perform calculations beyond the capacity of classical binary computing.

\ifSubfilesClassLoaded{%
  \nocite{apsrev41Control}%
  \bibliographystyle{apsrev4-1}%
  \bibliography{refs}%
}{}
\end{document}

\ifSubfilesClassLoaded{%
  \tableofcontents%
  \let\tableofcontents\relax%
  \setcounter{section}{5}%
}{}
\section{Quantum Benchmarking}
\label{benchmarking}

Benchmarking, the process of running a series of tests and measurements to obtain a quantitative performance metric for a system, is essential for assessing and comparing particular quantum devices, or understanding the broad development of QC, in terms of performance, fidelity, and scalability. Benchmarking can be leveraged to evaluate the quantum hardware, the algorithms, and the software stack. Depending on the purpose and granularity, from micro to macro, quantum benchmarking can be classified into four categories~\footnote{Note that different classifications for quantum benchmarks exist in the literature, e.g.~\cite{acuaviva2024benchmarking, tomesh2022supermarq, resch2021benchmarking}.}: (1) Microbenchmarking to characterize hardware features; (2) Randomized or synthetic benchmarks for measuring more generalized hardware performance; (3) Single application to measure single domain-specific performance metrics; (4) Comprehensive benchmark suites to provide more generalized, overall performance measure from multiple aspects. The obtained result can be a comprehensive score or certification to verifying the compliance of a quantum standard~\cite{eisert2020quantum,proctor2024benchmarking}. In the following, we provide a high-level overview of each category. \changes{Note that some techniques (e.g., randomized benchmarking) can be viewed as either microbenchmarking (to estimate average gate error rates) or synthetic benchmarking (as a device-level stress test), depending on how they are parameterized and what metric is reported.} For more in-depth introduction to quantum benchmarking, we refer the interested readers to e.g.~\cite{hashim2024practical, resch2021benchmarking}.

\subsection{Microbenchmarking}
\label{microbenchmarking}

The purpose of microbenchmarking is to characterize key performance features of a quantum device, such as coherence time, average gate fidelity, measurement/reset fidelity, etc. Quantum computers are impacted by noise~\cite{resch2021benchmarking}, which mainly originates from four different sources: (1)  Their interaction with external environment, which can be seen as an implicit measurement~\cite{barnes2017quantum}, as information leaks from the superposition/entangled state of interest. \changes{This is characterized by the energy-relaxation time $T_1$ (decay from $\ket{1}$ to $\ket{0}$) and the dephasing time $T_2$ (loss of relative phase coherence between $\ket{0}$ and $\ket{1}$), which together limit how long quantum information can be stored and processed reliably.} (2) Their unwanted interaction with neighboring qubits, known as crosstalk, which leads to a mixture of states or decoherence, and is hard to characterize; (3) Imperfect operations due to imperfect fabrication or calibration; (4) Leakage, which refers to the event where the state leaks to an unexpected state and leaves the encoded computational subspace. 

Microbenchmarks are designed to evaluate a quantum system by measuring the distance or distinguishability between two quantum states, typically involving the system's actual state and an ideal or target state. Quantum State Tomography (QST)~\cite{vogel1989determination} refers to the process of reconstructing the density matrix of a quantum state, which provides a complete description of the state, by performing a series of measurements on identically prepared quantum systems, with each measurement sampling a portion of the state through measurement projection. In practice, the estimated density matrix is often influenced by state preparation and measurement (SPAM) errors, which arise during qubit initialization and readout, and must be carefully accounted for in the QST process to get an accurate representation~\cite{SPAM_2021}.

Another goal of microbenchmarking is to determine what exact operations have been implemented by the hardware in the presence of unknown noise. Extending the idea of QST, Quantum Process Tomography (QPT)~\cite{weinstein2004quantum, o2004quantum} is designed to characterize and identify quantum operations, e.g., a particular gate or sequence of gates with noise, using a variety of known input states~\cite{chuang1997prescription, poyatos1997complete}. Much like QST, QPT is highly susceptible to SPAM errors since it relies on both accurate state preparation and measurement. Building on QPT, Quantum Gate Set Tomography (GST)~\cite{nielsen2021gate,greenbaum2015introduction} provides a complete characterization of gates, state preparation, and measurement, without assuming ideal conditions. By using "germ" sequences, GST amplifies systematic errors for precise measurement and directly addresses SPAM errors, making it effective for evaluating gate fidelity in noisy systems.

All three methods—QST, QPT and GST—have exponential resource overheads, making them computationally intensive as the number of qubits increases, with GST being the most demanding due to its robust SPAM error handling. These methods remain practical for a small number of qubits~\cite{li2023qasmbench}. To reduce such large overhead, randomized benchmarking~\cite{proctor2019direct, magesan2011scalable, magesan2012characterizing} is used to estimate the average error rates of individual quantum operations or small subsets of gates, offering a more efficient alternative as its resource requirements typically scale polynomially with the number of qubits. 
An alternative approach is to relax the objective of microbenchmarking by focusing on extracting partial information about the quantum state, such as its fidelity with a pure state, using techniques like Direct Fidelity Estimation~\cite{flammia2011direct}, which measures a constant number of Pauli observables and progressively guess which Pauli observables lead to discrepancies.

Microbenchmarking is generally performed by quantum device vendors or experimentalists to characterize a quantum device. While important, the low-level key performance features outlined at the start of this subsection are closely tied to the specific device under investigation and do not capture how the quantum hardware performs in practical, higher-level applications.

\subsection{Synthetic Benchmarking}
\label{synthetic benchmarking}

Compared to microbenchmarking, synthetic benchmarking evaluates the overall performance of a quantum computer by applying a sequence of randomly or synthetically generated quantum gates to a subset of qubits. This approach offers a more general and high-level performance metric, assessing how the system handles various gate operations and their associated errors. A single number is typically reported to reflect the overall computational power of a quantum device. The most widely recognized benchmark in this context is Quantum Volume (QV) ~\cite{bishop2017quantum, cross2019validating}, which captures multiple features such as gate fidelity, qubit connectivity, error rates, and reflects the combined effects of qubit number and result fidelity~\cite{pelofske2022quantum, baldwin2022re} 
QV determines the largest random circuit with equal width and depth that a QPU can successfully execute while generating correct outputs with a probability greater than 2/3~\cite{cross2019validating}. 

Additionally, volumetric benchmarking (VB), that plots performance in a 2d (depth x width) volumetric grid, was proposed by Sandia National Laboratories~\cite{blume2020volumetric} to address some limitations of QV, and recently the industrial consortium QED-C worked on extending it to a broader formalism~\cite{lubinski2024quantumalgorithmexplorationusing}. Another notable synthetic benchmark is RevLib~\cite{wille2008revlib}, which is a benchmark of synthesized reversible functions.

In this context, it is worth mentioning that a quantum counterpart to LINPACK~\cite{dongarra2003linpack}—the widely recognized benchmark for classical high-performance computing (HPC) systems—has been proposed~\cite{dong2021random}. This quantum benchmark evaluates a quantum computer's ability to solve random systems of linear equations using quantum singular value transformation (QSVT)~\cite{gilyen2019quantum}. To complement this approach, IBM introduced a metric analogous to FLOPS (floating-point operations per second)~\cite{dongarra2003linpack} for quantum systems, called circuit-layer operations per second (CLOPS), which quantifies runtime performance by measuring the execution of a series of Quantum Volume (QV) circuits~\cite{wack2021quality}.

Synthetic benchmarking provides a broad assessment of the performance of the quantum device under test. However, it is not tailored to specific algorithms or application contexts. These benchmarks often incorporate a degree of randomness, which has drawn criticism in the community. Some argue that real-world quantum applications may not rely on randomness or align with the specific synthetic patterns used in the benchmarking process, limiting its relevance to practical use cases~\cite{acuaviva2024benchmarking, tomesh2022supermarq, li2023qasmbench}.

\subsection{Application-specific Benchmarking}
\label{application specific benchmarking}

Benchmarks under this category are application-specific, focusing on real-world tasks or domain purposes, and typically more scalable. A simple example involves using a single qubit for quantum random number generation (QRNG)~\cite{herrero2017quantum}, to evaluate the bias and correlations in sequences typically produced by current NISQ devices~\cite{shikano2020detecting}. More complex NISQ examples include VQE with templated layers as the scaling factor for benchmarking~\cite{tomesh2022supermarq}, and Q-score which measures the largest MaxCut instance that quantum hardware can effectively solve via QAOA~\cite{van2022evaluating}.

FTQC benchmark examples include the Shor's algorithm for prime factorization~\cite{shor1999polynomial}, the Grover's algorithm for unstructured search~\cite{grover1996fast}, the HHL algorithm for solving quantum linear equations~\cite{harrow2009quantum}, etc. 

One important motivation for application-specific benchmarking is to explore and demonstrate two critical milestones in QC: quantum supremacy and quantum utility. Quantum supremacy refers to the ability of a quantum computer to solve a problem that is infeasible for classical computers. Google achieved this milestone by solving an artificial problem~\cite{arute2019quantum}. As previously outlined in section~\ref{use_cases_intro}, a more practically relevant milestone is quantum utility~\cite{kim2023evidence, herrmann2023quantum}, which focuses on the ability of a quantum computer to generate computational value that exceeds its cost~\cite{DARPA_QBI}, as an intermediate step towards a tangible quantum advantage for solving real-world problems.
In this context, Random Circuit Sampling (RCS) has emerged as a leading standard to identify where quantum computers might surpass classical supercomputers, even in the presence of noise~\cite{morvan2023phase,acharya2024quantum,RCS_Google}. It is based on a computational task believed to be intractable for classical supercomputers~\cite{arute2019quantum}. \changes{Because RCS is an artificial workload, it is best interpreted as a stress test for hardware and control at scale. Translating such demonstrations into application-level utility typically requires separate, domain-relevant benchmarks.}

\subsection{Benchmarking Suites}
\label{benchmarking suites}

These benchmarks aim to comprehensively evaluate a quantum device using a suite of quantum applications, similar to Spec~\cite{henning2006spec} and Parsec~\cite{bienia2008parsec} for classical sequential and parallel computing. Some initial efforts include QASMBench~\cite{li2023qasmbench}, SupermarQ~\cite{tomesh2022supermarq}, QED-C benchmarking~\cite{lubinski2024optimization}, MQTBench~\cite{quetschlich2023mqt}. Circuit features are defined ~\cite{li2023qasmbench, tomesh2022supermarq, bandic2024profiling, bandic2023interaction} to characterize these application benchmark suites.

The ultimate goal is to develop a standard benchmark suite to report a single or a number of metrics to symmetrically characterize the capability of a quantum device, and compare among each others, similar to report a CPU-Z score through the CPU-Z benchmarks for classical computing. IonQ introduced the concept of algorithmic qubits (\#AQ)~\cite{aqionq, chen2024benchmarking} by benchmarking six quantum application through VB~\cite{blume2020volumetric}. The \#AQ approach has been challenged by Quantinuum~\cite{aqquantinuum} as \#AQ results can be significantly enhanced through software methods, such as gate compilation and error mitigation with plurality voting. More general, comprehensive, and convincing benchmark suites as well as their metrics are still yet to mature and converge. Current major efforts in quantum benchmarking include the Quantum Benchmark Initiative (QBI) lead by DARPA~\cite{DARPA_QBI} and QED-C benchmarking~\cite{lubinski2024optimization} lead by NIST. 

Future challenges for quantum benchmarking include the investigation of appropriate benchmarks and metrics for different QEC codes~\cite{Fowler_2012, qeczoo} and heterogeneous QEC codes~\cite{stein2024architectures,wu2023enabling}, for quantum computation/memory separation~\cite{bluvstein2024logical,liu2023quantum}, and for hybrid~\cite{endo2021hybrid, mcclean2016theory}, distributed~\cite{haner2021distributed, wu2023qucomm}, and heterogeneous~\cite{stein2023hetarch} QC architectures.

\subsection{Joint HPC+QC benchmarking}
\label{joint benchmarking}
In the above discussion we mainly highlight tools that can be used to estimate and characterize the performance of quantum hardware. As HPC systems become more utilized in \textit{hybrid} HPC+QC algorithms, it will become important to also consider tools for benchmarking and profiling \textit{joint} HPC+QC workflows. \changes{A joint benchmark should report end-to-end workflow metrics (time-to-solution and, where possible, energy-to-solution) together with a disclosure of classical overheads (transpilation, mitigation, optimization, data movement, queueing) and the QC access mode.}

Here, the broader goal is to not just understand the core quantum performance, but also the often neglected accompanying classical computation, which may include the algorithm logic (e.g., the optimization in a variational-type application) or even pre- and post-processing steps (e.g., circuit transpilation, error mitigation related fitting, etc.). These classical computation tasks can often consume a non-trivial portion of the total runtime, especially as the systems, but also the applications, get larger and more capable. Crucially, tools that analyze these combined workflows will also be able to shed light on collective performance metrics that may not be captured when just-classical or just-quantum performance is studied in isolation. These may include latencies and communication bandwidth between the two hardware modalities, or even scheduler properties and performance, which could be important when many users require concurrent access to potentially limited resources.

Although not many such tools exist yet, some are being actively explored. One such example is QStone~\cite{qstoneGithub}, which is a python package collaboratively developed by Riverlane, ORNL and Rigetti. QStone aims to allow for benchmarking and profiling of hybrid HPC+QC workflow performance. Using simple json-based configuration files, a benchmark can specify details of the quantum hardware being used (e.g., qubit count, connection type, etc.), the number of hybrid HPC+QC users that need to be simulated, the type and workload of the applications that each of the simulated users should execute, and finally the type of scheduler that a given HPC system utilizes (naturally along with its relevant options). QStone then converts this configuration into benchmarking code that can be directly executed through the HPC scheduler. Once the jobs are ran, QStone organizes the resulting data into a data-frame that can be further analyzed or plotted. Currently a few basic applications are built-in (e.g., binary classifier, QEC decoding step, a Randomized Benchmarking protocol, etc.), but other, custom ones can be also easily integrated into use. Tools such as QStone will likely become more important in the longer term, as more HPC centers get quantum hardware integrated, and large-scale hybrid application workflows become widely utilized.


\subsection{Benchmarking different integration models}
\label{integration-level-benchmarking}
Benchmarking tight (co-located) versus loose (over-the-internet) HPC+QC integration needs to additionally compare integration-model-level metrics, not just quantum-device metrics. Beyond QPU fidelity or circuit latency, the comparison should include space and utilities: rack footprint, power and cooling capacity, and facility overhead captured via standardized KPIs such as Power Usage Efficiency (PUE)~\cite{ISO-PUE:2026} and Water Usage Efficiency (WUE)~\cite{ISO-WUE:2022}. Also operational performance (end-to-end time-to-solution, job throughput, queue/wait behavior, and retry/failure rates) should be included for a thorough comparison. 

Currently, there is very limited published data that standardizes comparisons across on-premises, tightly integrated deployments and provider-managed remote services. Additionally, widely accepted benchmarking methods that enable repeatable, ''apples-to-apples'' comparisons between these approaches have not yet been established. A practical path forward is to adopt conventional HPC data-center practices by pairing facility KPIs (PUE/WUE) with energy-efficiency and performance baselines already familiar in HPC, such as performance/Watt perspectives as popularized by Green500~\cite{Green500}, and suitable workload benchmarks. Complementing these efforts, the Quantum Energy Initiative (QEI) is working to develop shared methodologies and language for quantifying the physical resource and energy costs of quantum technologies, which could help normalize space-and-utilities comparisons across tight and loose integration models~\cite{QEI-article}.

It would also be useful to operationalize routine auditing of utilization, throughput, and service quality using tooling and reporting practices similar to XDMoD~\cite{XDMoD, Open-XDMoD}, which was designed to track utilization/performance and support operations planning at scale. For hybrid integration specifically, this suggests defining a small benchmark harness and disclosure checklist—interface mode (sync/async), batching strategy, reservation/queue state, metadata/telemetry availability, and repeating measurements over time windows to separate device behavior from network and provider/queue variability, analogous to how HPC communities standardize run rules and public reporting for comparability. 


\ifSubfilesClassLoaded{%
  \nocite{apsrev41Control}%
  \bibliographystyle{apsrev4-1}%
  \bibliography{refs}%
}{}
\end{document}

\ifSubfilesClassLoaded{%
  \tableofcontents%
  \let\tableofcontents\relax%
  \setcounter{section}{7}%
}{}
\section{Survey of ADAC members}
\label{survey}

\subsection{Survey Design}
\label{survey design}

The survey was conducted using a form with predefined questions asking for free-form responses from the perspective of the individual ADAC members. The questions were divided into three categories:

\begin{itemize}
    \item Current systems and plans related to QC: Do you have your own quantum computer(s) in place? If so, are they commercial systems, experimental tesbeds, or other? What qubit technologies are available? What are your future plans, e.g., timeline for acquiring (additional) quantum computers? Do you have ongoing collaboration with other HPC institutes regarding HPC+QC?
    \item User perspective: What are the most typical HPC use cases? Is there a demand for QC amongst the user base? What kind of feedback is coming from the users? Are there obstacles to the adoption of QC? Do you see a need for training, and do you provide training on HPC+QC?
    \item Future: Given that quantum computers keep improving, the user base in both academia and industry is expected to grow. How should HPC service providers deal with the increased demand? Do you plan to offer ready-made solutions or develop parts of the service in-house?
\end{itemize}

Next, we analyze the survey findings.

\subsection{Infrastructure and system integration}
\label{infrastructure}
The survey reveals a diverse and evolving landscape of QC infrastructure. While only a subset of institutions currently operate their own quantum computers, many are actively developing testbeds or accessing quantum systems through cloud platforms such as IBM Quantum and AWS Braket. The technologies in use include superconducting qubits, trapped ions, neutral atoms, and photonic systems.

Some institutions have begun integrating quantum systems into their HPC environments, either through hybrid platforms or orchestration tools that enable resource sharing. Others maintain separate operational domains for classical and quantum systems, citing differences in management, maturity, and user needs. Simulators and emulators are commonly used to support algorithm development and hybrid workflow testing, especially when physical quantum hardware is not yet available.

\subsection{User demand and application areas}
\label{user demand}

User interest in QC varies significantly between institutions. In some centers, demand is limited to a small group of researchers, while others report growing interest in fields such as quantum chemistry, materials science, optimization, and QML. Common HPC use cases include electronic structure calculations, fluid dynamics, AI/ML, and climate modeling.

Where demand exists, users are primarily interested in hybrid workflows, algorithm development, and feasibility studies using cloud-based quantum resources. However, the immaturity of quantum hardware and the lack of clear, demonstrable advantages over classical methods remain significant barriers to broader adoption.

\subsection{Training and adoption challenges}
\label{training}

A recurring theme across responses is the need for comprehensive training. Although introductory resources are widely available, there is a strong demand for advanced training focused on hybrid HPC+QC workflows, quantum information theory, and practical programming with quantum software development kits (SDKs). Some institutions have already launched public seminars and bootcamps, while others are in the process of developing more structured training programs.

Obstacles to adoption include limited access to quantum hardware, lack of skilled personnel, and the complexity of integrating quantum components into existing HPC workflows. Additionally, users often struggle with the experimental nature of current quantum systems and the absence of mature, user-friendly software stacks.

\subsection{Strategic planning and future directions}
\label{strategic planning}

Institutions anticipate a steady increase in demand for QC resources and prepare accordingly. Strategies include:

\begin{itemize}
    \item Developing middleware and orchestration tools for hybrid job scheduling.
    \item Establish testbeds for low-cost and low-latency experimentation.
    \item Enhancing interoperability between HPC and QC platforms.
    \item Collaborating on international funding initiatives to support infrastructure and training.
\end{itemize}

Some centers are focusing on customizing and deploying existing quantum frameworks, while others are investing in the development of new components tailored to their specific HPC environments. There is a shared recognition that no one-size-fits-all solution exists, and flexibility will be key to successful integration.

\subsection{Survey summary}
\label{survey summary}

The survey paints a picture of a field in transition. Although QC is not yet a mainstream component of HPC infrastructure, the groundwork is being laid across multiple fronts, from hardware testbeds and cloud access to training programs and hybrid software stacks. The path forward will require sustained investment, cross-institutional collaboration, and a commitment to building the technical and human infrastructure needed to realize the full potential of HPC+QC integration.

Despite growing interest and investment, the adoption of QC within HPC environments faces a range of significant challenges. One of the most frequently cited barriers is the immaturity of current quantum hardware. Many ADAC members report that today’s quantum systems are not yet capable of delivering practical advantages over classical HPC for most real-world applications. This technological gap is compounded by a lack of standardized software stacks and interoperability protocols, making integration with existing HPC workflows complex and resource-intensive.

Another major hurdle is the steep learning curve associated with quantum programming and quantum information theory. This skills gap is exacerbated by a shortage of specialized training programs, particularly those focused on hybrid HPC+QC workflows. Even where training is available, survey respondents note that it often remains at an introductory level. Additionally, access to quantum hardware can be limited, costly, or technically challenging. Finally, the absence of clear, demonstrable use cases that showcase quantum advantage in domains relevant to the broader HPC user base remains a critical obstacle. Without compelling success stories, many users and institutions remain hesitant to invest heavily in quantum technologies, preferring to wait for more mature, proven solutions.

\ifSubfilesClassLoaded{%
  \nocite{apsrev41Control}%
  \bibliographystyle{apsrev4-1}%
  \bibliography{refs}%
}{}
\end{document}

\ifSubfilesClassLoaded{%
  \tableofcontents%
  \let\tableofcontents\relax%
  \setcounter{section}{8}%
}{}
\section{Conclusions and discussion}
\label{conclusions}

The integration of QC with HPC represents both an extraordinary opportunity and a formidable challenge for the computational science community. As quantum technologies now mature from laboratory curiosities to potentially transformative computational tools, the HPC community must navigate a complex landscape of technical, practical, and strategic considerations.

QC capacity can be incorporated into HPC infrastructure in several different ways. Roughly, the integration can be divided into two main categories, loose cloud-like integration and tighter integration with classical data and compute infrastructure co-located with the QC hardware. In the longer term, QPUs may see even tighter coupling with binary computing units through on-chip integration. Presently, if implemented in the right way, both distributed and co-located solutions can bring quantum-enhancement to the fingertips of HPC end-users.

Quantum-accelerated HPC introduces new concepts and practices both for HPC infrastructure providers and end-users. On the operational side, the software stack connecting HPC and quantum infrastructure is still rather immature and constantly developing. No clear standards are in place. This makes straightforward deployment near-impossible, as the necessary HPC software stack is highly dependent on the given QC infrastructure to be integrated.

One of the main challenges for efficient co-processing of classical and quantum workloads within a single modeling workflow lies in co-scheduling of compute time on heterogeneous hardware. Where classical HPC resources, such as CPU and GPU capacity is relatively abundant, quantum capacity is scarce. Reconciling the target of high utilization grade of quantum resources with a smooth user experience is far from easy. A very diverse user-base with highly varying needs for utilizing the quantum resources significantly adds to the complexity. Further, it is difficult to a priori predict what kind of workloads end-users will want to deploy. We have outlined some of the most promising use cases here but note that we still do not know which of these will attract the largest user base. The potential profiles of quantum-enhanced applications for near-term quantum computers with sufficient capacity to generate widespread interest remain largely unexplored—only time will reveal their true impact.

From a user perspective, quantum computers behave very differently from traditional HPC. Due to the fundamental operating principles of quantum computers, 
\changes{individual circuit executions (``shots'') produce probabilistic outcomes. Statistical reproducibility is achieved by running many shots and reporting estimators (e.g., expectation values) together with uncertainty measures such as confidence intervals. In practice, the main reproducibility challenges are therefore not quantum mechanics per se, but hardware noise, device drift, recalibration, and changes in the quantum software stack or service environment.} In the near-term, quantum computers are also unreliable and error prone. This necessitates adopting a different mindset and cultivating an even greater level of healthy skepticism regarding the outcomes of a computation.

QC can also attract an entirely new user group for HPC centers, such as experimental physicists that see and use quantum computers as devices for physical experiments instead of computing machines. Making quantum-enhanced HPC more broadly applicable is a highly fruitful endeavor, but it does necessitate updated approaches to, for example, customer service and support.

The use of AI to improve the operating efficiency of quantum computers is one of the most pressing developments on the road to quantum utility. AI will play a critical role in handling various pre-processing and post-processing steps for tasks offloaded to quantum processors. The convergence of AI and QC will be crucial for managing increasing system complexity. 

Reliable benchmarking of combined HPC+AI+QC workflows will be important for a fair comparison of different implementations and combinations of hardware and software components. Benchmarking needs to extend from microbenchmarking close to quantum hardware level to full, real-world application level of workflows combining HPC, AI, and QC.

The ADAC community survey revealed a divided potential user base. On the one hand, there is strong interest from early adopters eager to experiment with the new possibilities offered by QC. A significant group of existing end users comprises developers who create solutions for the complete software stack required for both quantum and AI-accelerated HPC, encompassing quantum software development as well as classical software solutions. On the other hand, a sizable portion of the user community, including some HPC centers, remains hesitant due to the relatively immature technology readiness level of QC. This group is waiting for quantum-enhanced HPC to demonstrate a clear advantage over traditional classical computing approaches.

QC continues to mature, both on the hardware and software fronts. Estimates regarding when quantum computing will deliver tangible benefits for end-users, who are generally less concerned with the underlying compute technology but more focused on solving specific modeling workflows, vary significantly. \changes{It is arguably more fruitful to frame progress in terms of milestones rather than specific calendar years: (i) near-term, workload-specific quantum utility on NISQ-era devices in hybrid workflows; (ii) early fault-tolerant demonstrations with small numbers of logical qubits and measurable end-to-end workflow value; and (iii) domain-relevant quantum advantage at scale for scientifically or industrially important workloads. The timeline for these milestones depends strongly on improvements in qubit quality, reductions in QEC overhead, quantum algorithm development, and the maturity of hybrid software, orchestration, and operational practices.}

Already now, quantum computers enable basic research that could not be performed with other means. This is presently restricted to so-called non-computing applications, where quantum computers are used directly for physical experiments instead of solving mathematical equations. It is worth noting that also here, classical and AI-enhanced pre- and post-processing are crucial to get the most out of this nascent technology. For instance, enhancing the signal-to-noise ratios of physical experiments often demands substantial classical HPC capacity. In this context, AI-assisted methods are emerging as essential tools of the trade.

AI-assisted QC is regarded as a crucial element of quantum-accelerated HPC. With the increased complexity brought about by increasing qubit counts in relation to, e.g., compilation, hardware-aware optimization, and error mitigation and correction, brute-force classical methods become too resource-consuming without the speed-up provided by AI. The synergy loop between AI and QC provides exciting prospects for the future of HPC.

\changes{From an HPC-center perspective, already two near-term actions have high leverage. First, invest in implementing a hybrid workflow infrastructure: orchestration, scheduling interfaces, reproducibility metadata capture, and user-facing tooling that makes QPU access look like a managed accelerator service. For this, a co-located QPU is not necessary, even a QPU emulator can be sufficient. Second, build training pathways for both users and operations staff that cover quantum computing fundamentals, hybrid software development, and the emerging role of AI tools for enhancing QC. Focus on practical skills so that adoption is not limited to a few specialists only.}

Summarizing, integration of QC with HPC is now in full force around many HPC centers around the world. Undeniably, the full potential of QC for traditional scientific computing or industrial use cases still awaits unlocking. Recent advancements in both quantum hardware and software suggest that 
\changes{QC} is becoming relevant to traditional HPC users at a faster pace than previously anticipated. For HPC centers looking to future-proof their service offerings for their customers, engaging \changes{with}, or at least exploring QC would be prudent.

\ifSubfilesClassLoaded{%
  \nocite{apsrev41Control}%
  \bibliographystyle{apsrev4-1}%
  \bibliography{refs}%
}{}
\end{document}

\section*{Acknowledgment}

The authors acknowledge the support of the entire ADAC community, and especially those who responded to the questionnaire.

This research used resources of the Oak Ridge Leadership Computing Facility, which is a DOE Office of Science User Facility supported under Contract DE-AC05-00OR22725.

This research was supported by the Laboratory Directed Research and Development (LDRD) Program at Pacific Northwest National Laboratory (PNNL). PNNL is a multi-program national laboratory operated for the U.S. Department of Energy (DOE) by Battelle Memorial Institute under Contract No. DE-AC05-76RL01830.

This research used resources of the National Energy Research Scientific Computing Center (NERSC), a U.S. Department of Energy Office of Science User Facility located at Lawrence Berkeley National Laboratory, operated under Contract No. DE-AC02-05CH11231.

This research used resources of the National Institute of Advanced Industrial Science and Technology (AIST), which was done for Council for Science, Technology and Innovation (CSTI), Cross-ministerial Strategic Innovation Promotion Program (SIP), “Promoting the application of advanced quantum technology platforms to social issues” (Funding agency : QST).

This research was supported by the JHPC-quantum and JPNP20017 projects commissioned by the New Energy and Industrial Technology Development Organization (NEDO), a national research and development agency in Japan.


This research used resources of the Centre for Development of Advanced Computing (C-DAC) which is a Scientific Society under Ministry of Electronics and Information Technology (MeitY), Government of India.




\bibliographystyle{elsarticle-num}
\bibliography{refs} 

\end{document}